\documentclass[a4paper,12pt]{article}

\usepackage{graphicx, amsmath, amsfonts, amssymb, array, natbib, graphicx, float, lipsum, tikz, lscape}

\tikzset{
  treenode/.style = {shape=rectangle,
                     draw, align=center,
                     top color=white, text height = 0.5cm},
  root/.style     = {treenode, font=\Large},
  env/.style      = {treenode, font=\Large},
  dummy/.style    = {shape=circle,draw,align=center, top color=white, font=\large}
}

\title{Bayesian additive regression trees and the General BART model}
\author{Yaoyuan Vincent Tan and Jason Roy}

\begin{document}
\maketitle
\linespread{2}
\selectfont

\begin{abstract}
Bayesian additive regression trees (BART) is a flexible prediction model/machine learning approach that has gained widespread popularity in recent years. As BART becomes more mainstream, there is an increased need for a paper that walks readers through the details of BART, from what it is to why it works. This tutorial is aimed at providing such a resource. In addition to explaining the different components of BART using simple examples, we also discuss a framework, the General BART model, that unifies some of the recent BART extensions, including semiparametric models, correlated outcomes, statistical matching problems in surveys, and models with weaker distributional assumptions. By showing how these models fit into a single framework, we hope to demonstrate a simple way of applying BART to research problems that go beyond the original independent continuous or binary outcomes framework.
\end{abstract}


\section{Introduction}
Owing to its success, Bayesian additive regression trees \citep[BART;][]{chipman_bart} has gained popularity in the recent years among the research community with numerous applications including biomarker discovery in proteomic studies \citep{hernandez}, estimating indoor radon concentrations \citep{kropat}, estimation of causal effects \citep{leonti, hill}, genomic studies \citep{liu}, hospital performance evaluation  \citep{liu_hospital}, prediction of credit risk \citep{zhang_hardle}, predicting power outages during hurricane events \citep{nateghi}, prediction of trip durations in transportation \citep{chipman_transport}, and somatic prediction in tumor experiments \citep{ding_bioinf}. BART has also been extended to survival outcomes \citep{bonato, sparapani}, multinomial outcomes \citep{kindo, agarwal}, and semi-continuous outcomes \citep{linero_shared}. In the causal inference literature, notable papers that promote the use of BART include \cite{hill} and \cite{green}. BART has also been consistently among the best performing methods in the Atlantic causal inference data analysis challenge \citep{hill_2,hahn,dorie}. In addition, BART has been making inroads in the missing data literature. For the imputation of missing covariates, \cite{xu} proposed a way to utilize BART for the sequential imputation of missing covariates, while \cite{kapelner_miss} proposed to treat missingness in covariates as a category and set up the splitting criteria so that the eventual likelihood in the Metropolis-Hasting (MH) step of BART is maximized. For the imputation of missing outcomes, \cite{tan_miss} examined how BART can improve the robustness of existing doubly robust methods in situations where it is likely that both the mean and propensity models could be misspecified. Other more recent attempts to utilize or extend BART include applying BART to quantile regression \citep{kindo2}, extending BART to count responses \citep{murray}, using BART in functional data \citep{starling}, applying BART to recurrent events \citep{sparapani_rec}, identifying subgroups using BART \citep{sivaganesan,schnell,schnell2}, and using BART as a robust model to impute missing principal strata to account for selection bias due to death \citep{tan_cbd}. 

The widespread use of BART has resulted in many researchers starting to use BART as the reference model for comparison when proposing new statistical or prediction methods which are flexible and/or robust to model misspecification. A few recent examples include \cite{liang}, \cite{nalenz}, and \cite{lu}. This growing interest for BART raises a need for an in-depth tutorial paper on this topic to help researchers who are interested in using BART better understand the method that they are using and possibly diagnose the likely problems when unexpected results occur. The first portion of this paper is aimed at addressing this.

The second portion of our work revolves around an interesting observation on four works extending BART beyond the original independent continuous or binary outcomes setup. In these papers, they extend BART to semiparametric situations \citep{roy_semibart}, correlated outcomes \citep{tan_ribart}, survey \citep{zhang}, and robust error assumptions \citep{george}. Although these papers were written separately, they surprisingly share a common feature in their framework. In brief, when estimating the posterior distribution, they subtract a latent variable from the outcome and then model this residual as BART. This idea, although simple, is powerful because this can allow researchers to easily extend BART to problems that they may face in their dataset without having to rewrite or re-derive the Monte Carlo Markov Chain (MCMC) procedure for drawing the regression trees in BART. We summarize this idea in a framework unifying these models that we call the General BART model. We suggest how the priors could be set and how the posterior distribution could be estimated. We then show how General BART is related to the these four models. We believe that by presenting our General BART model framework and linking it with the models in these four papers as examples, it will aid researchers who are  trying to incorporate and extend BART to solve their research problems.

Our in-depth review of BART in Section \ref{review} focuses on three commonly asked questions regarding BART: What gives BART flexibility? Why is it called a sum of regression trees? What are the mechanics of the BART algorithm? In Section \ref{bart_app}, we demonstrate the superior performance of BART compared to the Bayesian linear regression (BLR) when data are generated from a complicated model.  We then describe the application of  BART to two real-life datasets, one with continuous outcomes and the other with binary outcomes. Section \ref{gen_bart} lays out the framework for our General BART model that allows BART to be extended to semiparametric models, correlated outcomes, survey, and situations where a more robust assumption for the error term is needed. We then show how our General BART model is related to these four BART extension models. In each of these description examples, we describe how the prior distributions are set and how the posterior distribution is obtained. We conclude with a discussion in Section \ref{conclude}.

\section{Bayesian additive regression trees}
\label{review}
We begin our discussion with the independent continuous outcomes BART because this is the most natural way to explain BART. We argue that BART is flexible because it is able to handle non-linear main effects and multi-way interactions without much input from researchers. To demonstrate how BART handles these model features, we explain using a visual example of a regression tree. We then illustrate the concept of a sum of regression trees using a simple example with two regression trees. We next show how a sum of regression trees link with non-linearity. To show how BART determines these non-linear main and multi-way interaction effects automatically, we discuss two perspectives. First, we provide a visual and detailed breakdown of the BART algorithm at work using a simple example, providing intuition for each step along the way. Then, we provide a more rigorous explanation of the BART MCMC algorithm by discussing the prior distribution used for BART and how the posterior distribution is calculated. Finally, we show how these ideas can be extended to independent binary outcomes.
 
\subsection{Continuous outcomes}
\subsubsection{Formal definition}
We begin with the formal definition and notation of BART. Suppose we have a continuous outcome $Y$ and $p$ covariates $X$ for $n$ subjects. The goal is a model that can capture complex relationships between $X$ and $Y$, with the aim of using it for prediction. BART attempts to estimate $f(x)$ from models of the form $Y=f(X)+\varepsilon_i$, where, for now, $\varepsilon_i\sim N(0,\sigma^2)$, $i=1,\cdots, n$. To estimate $f(X)$, a sum of regression trees is specified as
\begin{equation}
	\label{bart_con}
	f(X)=\sum_{j=1}^m g(X;T_j,M_j).
\end{equation}
 In Equation (\ref{bart_con}), $T_j$ is the $j^{\text{th}}$ binary tree structure and $M_j=\{\mu_{1j},\ldots,\mu_{b_{jj}}\}$ is the vector of terminal node parameters associated with $T_j$. Note that $T_j$ contains the information of which covariate to split on,  the cutoff value in an internal node, as well as where the internal node is located in the binary tree. The constant $m$ is usually  fixed at a large number, e.g.,  200. 
 
We will next make much more clear what is meant by $T_j$ and $M_j$, and also how this leads to extremely flexible models. We begin with the simple case of a single regression tree.

\subsubsection{Single regression tree}
To understand BART, consider first a single regression tree, rather than a sum of trees. For now we will assume that the tree is known and just focus on how to interpret it and obtain predictions from it. Later, we will describe the priors on these trees since the true tree structure is usually unknown. 

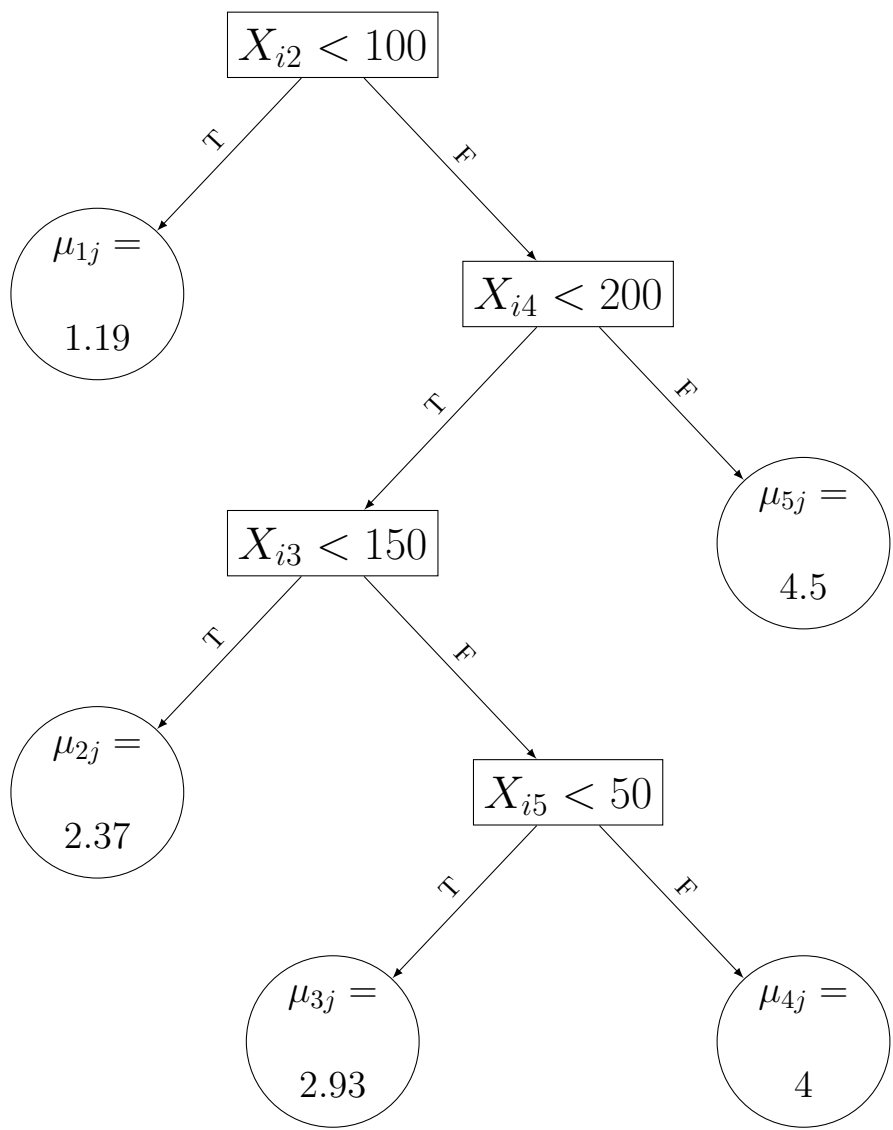
\begin{figure}[H]
\caption{Example of a regression tree $g(X;T_j,{M}_j)$ where $\mu_{kj}$ is the mean parameter of the $k^{\text{th}}$ node for the $j^{\text{th}}$ regression tree. \label{tree_ex}}
	\vspace*{0.2cm}
	\centering
	\begin{tikzpicture}
  		[
   	 	grow                    = down,
    	sibling distance        = 15em,
    	level distance          = 8em,
    	edge from parent/.style = {draw, -latex},
    	every node/.style       = {font=\footnotesize}, sloped
  		]
  		\node [root] {$X_{i2}<100$}
    		child { node [dummy] {$\mu_{1j}=$\\$1.19$}
      edge from parent node [above] {T} }
    child { node [env] {$X_{i4}<200$} child {node [env] {$X_{i3}<150$} child { node [dummy] {$\mu_{2j}=$\\$2.37$} edge from parent node [above] {T}} child { node [env] {$X_{i5}<50$} child { node [dummy] {$\mu_{3j}=$\\$2.93$} edge from parent node [above] {T} } child { node [dummy] {$\mu_{4j}=$\\$4$} edge from parent node [above] {F}} edge from parent node [above] {F}} edge from parent node [above] {T}} child {node [dummy] {$\mu_{5j}=$\\$4.5$} edge from parent node [above] {F}} edge from parent node [above] {F} };
	\end{tikzpicture}
\end{figure}

Consider the regression tree $g(X;T_j,M_j)$ given in Figure \ref{tree_ex}.
Imagine that we have covariates $X_i=(X_{i1},\ldots,X_{i5})$ and we would like to know $E(Y_i|X_i)$ for subject $i$. Each place where there is a split is called a node. At the top node ({\em root}), there is a condition $X_{i2}<100$. If $X_{i2}<100$ is true, then we follow the path to the left, otherwise to the   right. Assuming that $X_{i2}<100$ is true, we see that we arrive at a node which is not split upon. This is called a {\em terminal node} and the parameter $\mu_{1j}=1.19$ would be used as the predicted value of $Y_i$.  Suppose instead that $X_{i2}<100$ is {\em{not}} true. Then, moving along the right side, another internal node with condition $X_{i4}<200$ is encountered. This condition would be checked and, if this condition is true (false), we  would follow the path to the left (right).  This process continues until we reach a terminal node and the parameter $\mu_{kj}$ in that terminal node is assigned as the predicted value of $Y_i$. Note that $\mu_{kj}$ is the mean parameter of the $k^{\text{th}}$ node for the $j^{\text{th}}$ regression tree. So, for example, a subject $i$ with $X_{i1}=30, X_{i2}=120, X_{i3}=115, X_{i4}=191$, and $X_{i5}=56$ would be assigned a predicted outcome of $\mu_{2j}=2.37$. The prediction would be exactly the same for another subject $i'$ who instead had covariates $X_{i'1}=130, X_{i'2}=135, X_{i'3}=92, X_{i'4}=183, X_{i'5}=10$.

In summary, we can view a regression tree as a function that assigns the conditional mean of $Y_i$ to the parameter $\mu_{kj}$ i.e. $\mu_{kj}=g(X_i;T_j,M_j)\mapsto E(Y_i|X_i)$. Note that we have not yet discussed how a tree is created and how uncertainty about what to split on and where to split is quantified. We will address that when we introduce priors and algorithms.

{\bf{Regression tree as an analysis of variance (ANOVA) model.}}
Another way to think of the regression tree in Figure \ref{tree_ex} is to view it as the following analysis of variance (ANOVA) model:
\begin{align*}
	Y_i&=\mu_{1j}I\{X_{i2}<100\}+\mu_{2j}I\{X_{i2}\geq 100\}I\{X_{i4}<200\}I\{X_{i3}<150\}\\
	&\quad+\mu_{3j}I\{X_{i2}\geq 100\}I\{X_{i4}<200\}I\{X_{i3}\geq 150\}I\{X_{i5}<50\}\\
	&\quad+\mu_{4j}I\{X_{i2}\geq 100\}I\{X_{i4}<200\}I\{X_{i3}\geq 150\}I\{X_{i5}\geq 50\}\\
	&\quad+\mu_{5j}I\{X_{i2}\geq 100\}I\{X_{i4}\geq 200\}+\varepsilon_i,
\end{align*}
where $I\{.\}$ is the indicator function and $\varepsilon_i\sim N(0,\sigma^2)$. We can see that the term $\mu_{1j}I\{X_{i2}<100\}$ corresponds to the terminal node on the top left corner of Figure \ref{tree_ex}, $\mu_{2j}I\{X_{i2}\geq 100\}I\{X_{i4}<200\}I\{X_{i3}<150\}$ correspond to the terminal node on the middle right of Figure \ref{tree_ex}, and so on. We can think of $\mu_{1j}I\{X_{i2}<100\}$ as a main effect, because it only involves the second variable $X_{i2}$, while $\mu_{2j}I\{X_{i2}\geq 100\}I\{X_{i4}<200\}I\{X_{i3}<150\}$ is a three way interaction effect involving the second ($X_{i2}$), fourth ($X_{i4}$), and third variable ($X_{i3}$). By viewing a regression tree as an ANOVA model, we can easily see why a regression tree and hence, BART, is able to handle main and multi-way interaction effects.

\subsubsection{Sum of regression trees}
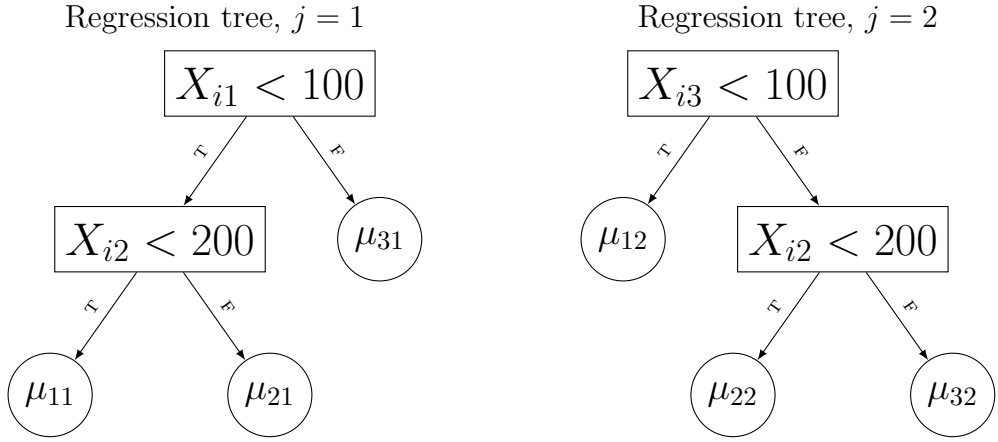
\begin{figure}[H]
	\caption{Illustrating the sum of regression trees using a simple two regression tree example.\label{tree}}
	\centering
	\begin{tabular}{ccc}
		Regression tree, $j=1$ & $\quad\quad\quad$ & Regression tree, $j=2$ \\
		\begin{tikzpicture}
  		[
   	 	grow                    = down,
    	sibling distance        = 7em,
    	level distance          = 5em,
    	edge from parent/.style = {draw, -latex},
    	every node/.style       = {font=\tiny}, sloped
  		]
  		\node [root] {$X_{i1}<100$} child {node [env] {$X_{i2}<200$} child {node [dummy] {$\mu_{11}$}
      edge from parent node [above] {T}} child {node [dummy] {$\mu_{21}$}
      edge from parent node [above] {F}} edge from parent node [above] {T}}
    		child { node [dummy] {$\mu_{31}$}
      edge from parent node [above] {F}};
		\end{tikzpicture} & $\quad\quad\quad$ &
		\begin{tikzpicture}
  		[
   	 	grow                    = down,
    	sibling distance        = 7em,
    	level distance          = 5em,
    	edge from parent/.style = {draw, -latex},
    	every node/.style       = {font=\tiny}, sloped
  		]
  		\node [root] {$X_{i3}<100$} child { node [dummy] {$\mu_{12}$}
      edge from parent node [above] {T}}
    		child { node [env] {$X_{i2}<200$} child {node [dummy] {$\mu_{22}$}
      edge from parent node [above] {T}} child {node [dummy] {$\mu_{32}$}
      edge from parent node [above] {F}} edge from parent node [above] {F}};
		\end{tikzpicture} \\
	\end{tabular}		
\end{figure}

We next consider a sum of regression trees. To illustrate the main idea, we  focus on an example with $m=2$ trees and $p=3$ covariates. Suppose we were given the two trees in Figure \ref{tree}. 

The resulting conditional mean of $Y$ given $X$ is $\sum_{j=1}^2g(X;T_j,M_j)$. Consider the hypothetical data from $n=10$ subjects given in Table \ref{tree_ex_post}. We can see that the quantity that is being `summed' and eventually allocated to $E(Y_i|X_i)$ is not the regression tree or tree structure, but, the value that each $j^{\text{th}}$ tree structure assigns to subject $i$. This is one way to think of a sum of regression trees. It allocates a sum of parameters $\mu_{kj}$ to subject $i$. Note that contrary to initial intuition, it is the sum of $\mu_{kj}$ that is allocated to rather than the mean of the $\mu_{kj}$'s. This is mainly because BART calculates each posterior draw of the regression tree function $g(X;T_j,M_j)$ using a leave-one-out concept, which we shall elaborate on shortly. 

\begin{table}[H]
\caption{The values of $\sum_{j=1}^2g(X;T_j,M_j)$ from the regression trees in Figure \ref{tree}. \label{tree_ex_post}}
\begin{center}
	\begin{tabular}{|c|c|c|c|c|c|c|c|}
		\hline
		$i$ & $Y$ &  $X_1$ & $X_2$ & $X_3$ & $g(X;T_1,M_1)$ & $g(X;T_2,M_2)$ & $\sum_{j=1}^2g(X;T_j,M_j)$ \\
		\hline
		1 & $Y_1$ & -182 & 235 & -333 & $\mu_{21}$ & $\mu_{12}$ & $\mu_{21}+\mu_{12}$ \\
		2 & $Y_2$ & 54 & 339 & 244& $\mu_{21}$ & $\mu_{22}$ & $\mu_{21}+\mu_{22}$ \\
		3 & $Y_3$ & -106 & -50 & -682& $\mu_{11}$ & $\mu_{12}$ & $\mu_{11}+\mu_{12}$ \\
 		4 & $Y_4$ & -80 & -62 & -320 & $\mu_{11}$ & $\mu_{12}$ & $\mu_{11}+\mu_{12}$ \\
 		5 & $Y_5$ & -123 & 198 & -77 & $\mu_{11}$ & $\mu_{12}$ & $\mu_{11}+\mu_{12}$ \\
		6 & $Y_6$ & 175 & 108 & -46 & $\mu_{31}$ & $\mu_{12}$ & $\mu_{31}+\mu_{12}$ \\
		7 & $Y_7$ & -44 & 11 & 136& $\mu_{11}$ & $\mu_{22}$ & $\mu_{11}+\mu_{22}$ \\
		8 & $Y_8$ & -131 & -10 & -70& $\mu_{11}$ & $\mu_{12}$ & $\mu_{11}+\mu_{12}$ \\
		9 & $Y_9$ & -56 & 68 & 257 & $\mu_{11}$ & $\mu_{22}$ & $\mu_{11}+\mu_{22}$ \\
		10 & $Y_{10}$ & 7 & 324 & 282& $\mu_{21}$ & $\mu_{32}$ & $\mu_{21}+\mu_{32}$ \\
		\hline
	\end{tabular}
\end{center}
\end{table}

Another way to view the concept of a sum of regression trees is to think of the regression trees in Figure \ref{tree} as ANOVA models. Then, the sum of trees is the following ANOVA model:
\begin{align*}
	Y_i&=g(X;T_1,M_1)+g(X;T_2,M_2)+\varepsilon_i \nonumber\\
	&=\mu_{11}I\{X_{i1}<100\}I\{X_{i1}<200\}+\mu_{21}I\{X_{i1}<100\}I\{X_{i1}\geq 200\}+\mu_{31}I\{X_{i1}\geq 100\}\nonumber\\
	&+\mu_{12}I\{X_{i3}<100\}+\mu_{22}I\{X_{i3}\geq 100\}I\{X_{i2}<200\}\nonumber\\
	&+\mu_{32}I\{X_{i3}\geq 100\}I\{X_{i2}\geq 200\}+\varepsilon_i.
\end{align*}

{\bf{Non-linearity of BART.}}
From this simple example, we can see how BART handles non-linearity. Each single regression tree is a simple step-wise  function or ANOVA model. When we sum regression trees together, we are actually summing together these ANOVA models or step-wise functions, and, as a result, we eventually obtain a more complicated step-wise function which can approximate the non-linearities in the main and multiple-way interactions. It is this ability to handle non-linear main and multiple-way interaction effects that makes BART a flexible model. But unlike many flexible models, BART does not require the researcher to specify the main and multi-way interaction effects.

{\bf{Prior distributions.}}
In the examples above, we have taken the trees as a given, including which variables to split on, the splitting values, and the mean parameters at each terminal node. In practice, each $g(X;T_j,M_j)$ is unknown. We therefore need  prior distributions for these functions. Thus, we can also think of BART as a Bayesian model where the mean function itself is unknown. A major advantage of this approach is that uncertainty about both the functional form and the parameters will be accounted for in the posterior predictive distribution of $Y$. 

Before getting into the details of the prior distributions and MCMC algorithm, we will first walk through a simple example to build the intuition.

\subsubsection{BART machinery: a visual perspective}
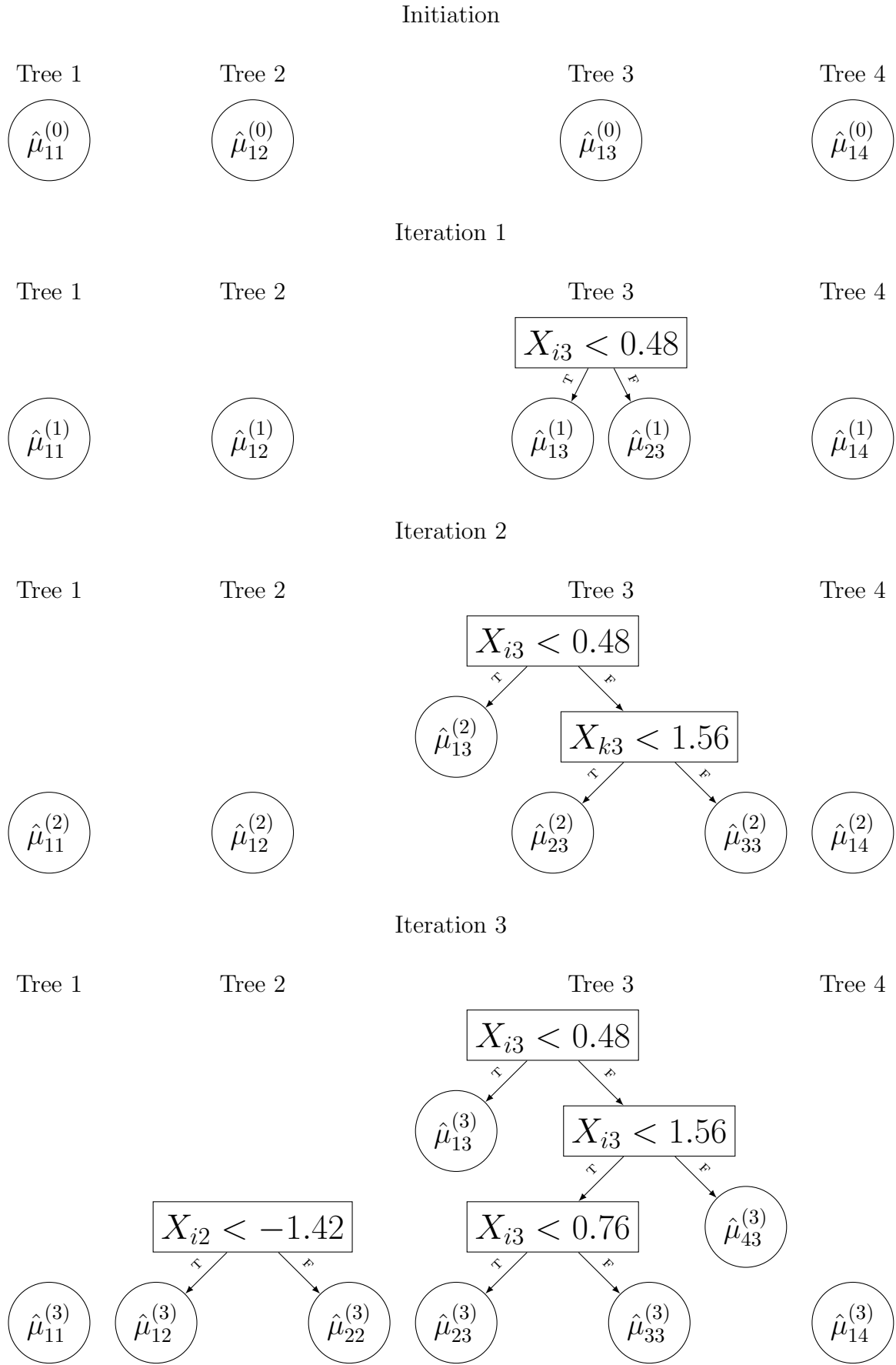
\begin{figure}[H]
\caption{Intuition of BART to iteration 3 of the MCMC steps within BART with $m=4$. \label{bart_algo_mcmc13}}
\centering
	\begin{tabular}{cccc}
		\multicolumn{4}{c}{Initiation} \\
		Tree 1 & Tree 2 & Tree 3 & Tree 4 \\
		\begin{tikzpicture}
  		[
   	 	every node/.style       = {font=\tiny}, sloped
  		]
  		\node [dummy] {$\hat{\mu}_{11}^{(0)}$};
		\end{tikzpicture} &
		\begin{tikzpicture}
  		[
   	 	every node/.style       = {font=\tiny}, sloped
  		]
  		\node [dummy] {$\hat{\mu}_{12}^{(0)}$};
		\end{tikzpicture} &
		\begin{tikzpicture}
  		[
   	 	every node/.style       = {font=\tiny}, sloped
  		]
  		\node [dummy] {$\hat{\mu}_{13}^{(0)}$};
		\end{tikzpicture} &
		\begin{tikzpicture}
  		[
   	 	every node/.style       = {font=\tiny}, sloped
  		]
  		\node [dummy] {$\hat{\mu}_{14}^{(0)}$};
		\end{tikzpicture}\\
		\multicolumn{4}{c}{Iteration 1} \\
		Tree 1 & Tree 2 & Tree 3 & Tree 4 \\
		\begin{tikzpicture}
  		[
   	 	every node/.style       = {font=\tiny}, sloped
  		]
  		\node [dummy] {$\hat{\mu}_{11}^{(1)}$};
		\end{tikzpicture} &
		\begin{tikzpicture}
  		[
   	 	every node/.style       = {font=\tiny}, sloped
  		]
  		\node [dummy] {$\hat{\mu}_{12}^{(1)}$};
		\end{tikzpicture} &
		 \begin{tikzpicture}
  		[
   	 	grow                    = down,
    	sibling distance        = 4em,
    	level distance          = 4em,
    	edge from parent/.style = {draw, -latex},
    	every node/.style       = {font=\tiny}, sloped
  		]
  		\node [root] {$X_{i3}<0.48$} child { node [dummy] {$\hat{\mu}_{13}^{(1)}$} edge from parent node [above] {T}}
    		child { node [dummy] {$\hat{\mu}_{23}^{(1)}$} edge from parent node [above] {F}};
		\end{tikzpicture} &
		\begin{tikzpicture}
  		[
   	 	every node/.style       = {font=\tiny}, sloped
  		]
  		\node [dummy] {$\hat{\mu}_{14}^{(1)}$};
		\end{tikzpicture} \\
		\multicolumn{4}{c}{Iteration 2} \\
		Tree 1 & Tree 2 & Tree 3 & Tree 4 \\
		\begin{tikzpicture}
  		[
   	 	every node/.style       = {font=\tiny}, sloped
  		]
  		\node [dummy] {$\hat{\mu}_{11}^{(2)}$};
		\end{tikzpicture} &
		\begin{tikzpicture}
  		[
   	 	every node/.style       = {font=\tiny}, sloped
  		]
  		\node [dummy] {$\hat{\mu}_{12}^{(2)}$};
		\end{tikzpicture} &
		 \begin{tikzpicture}
  		[
   	 	grow                    = down,
    	sibling distance        = 8em,
    	level distance          = 4em,
    	edge from parent/.style = {draw, -latex},
    	every node/.style       = {font=\tiny}, sloped
  		]
  		\node [root] {$X_{i3}<0.48$} child {node [dummy] {$\hat{\mu}_{13}^{(2)}$}edge from parent node [above] {T}}
    		child {node [env] {$X_{k3}<1.56$} child {node [dummy] {$\hat{\mu}_{23}^{(2)}$} edge from parent node [above] {T}} child {node [dummy] {$\hat{\mu}_{33}^{(2)}$} edge from parent node [above] {F}} edge from parent node [above] {F}};
		\end{tikzpicture} &
		\begin{tikzpicture}
  		[
   	 	every node/.style       = {font=\tiny}, sloped
  		]
  		\node [dummy] {$\hat{\mu}_{14}^{(2)}$};
		\end{tikzpicture} \\
		\multicolumn{4}{c}{Iteration 3} \\
		Tree 1 & Tree 2 & Tree 3 & Tree 4 \\
		\begin{tikzpicture}
  		[
   	 	every node/.style       = {font=\tiny}, sloped
  		]
  		\node [dummy] {$\hat{\mu}_{11}^{(3)}$};
		\end{tikzpicture} &
		\begin{tikzpicture}
  		[
   	 	grow                    = down,
    	sibling distance        = 8em,
    	level distance          = 4em,
    	edge from parent/.style = {draw, -latex},
    	every node/.style       = {font=\tiny}, sloped
  		]
  		\node [root] {$X_{i2}<-1.42$} child{node [dummy] {$\hat{\mu}_{12}^{(3)}$} edge from parent node [above] {T}} child{node [dummy] {$\hat{\mu}_{22}^{(3)}$} edge from parent node [above] {F}};
		\end{tikzpicture} &
		 \begin{tikzpicture}
  		[
   	 	grow                    = down,
    	sibling distance        = 8em,
    	level distance          = 4em,
    	edge from parent/.style = {draw, -latex},
    	every node/.style       = {font=\tiny}, sloped
  		]
  		\node [root] {$X_{i3}<0.48$} child {node [dummy] {$\hat{\mu}_{13}^{(3)}$}edge from parent node [above] {T}}
    		child {node [env] {$X_{i3}<1.56$} child {node [env] {$X_{i3}<0.76$} child{ node [dummy] {$\hat{\mu}_{23}^{(3)}$} edge from parent node [above] {T}} child{ node [dummy] {$\hat{\mu}_{33}^{(3)}$} edge from parent node [above] {F}} edge from parent node [above] {T}} child {node [dummy] {$\hat{\mu}_{43}^{(3)}$} edge from parent node [above] {F}} edge from parent node [above] {F}};
		\end{tikzpicture} &
		\begin{tikzpicture}
  		[
   	 	every node/.style       = {font=\tiny}, sloped
  		]
  		\node [dummy] {$\hat{\mu}_{14}^{(3)}$};
		\end{tikzpicture} \\
	\end{tabular}
\end{figure}

In our simple example, we have three covariates $X=(X_1,X_2,X_3)$ and a continuous outcome $Y$. We run the BART MCMC with four regression trees for 5 iterations on this dataset and at each iteration, we present the regression tree structures to illustrate how the BART machinery works as it goes through each MCMC step. When $Y$ and $X$ are provided to BART, BART first initializes the four regression trees to single root nodes (See ``Initiation'' in Figure \ref{bart_algo_mcmc13}). Since all four regression trees are single root nodes, the parameters initialized for these nodes would be $\mu_{ij}=\frac{\bar{{Y}}}{m}=\frac{\bar{{Y}}}{4}$. 

\begin{landscape}
\vspace*{-2cm}
\begin{figure}[H]
\caption{Iterations 4 and 5 of the MCMC steps within BART with $m=4$. \label{bart_algo_mcmc45}}
	\centering
	\begin{tabular}{cccc}
		\multicolumn{4}{c}{Iteration 4} \\
		Tree 1 & Tree 2 & Tree 3 & Tree 4 \\
		\begin{tikzpicture}
  		[
   	 	grow                    = down,
    	sibling distance        = 8em,
    	level distance          = 4em,
    	edge from parent/.style = {draw, -latex},
    	every node/.style       = {font=\tiny}, sloped
  		]
  		\node [root] {$X_{i2}<-0.58$} child{node [dummy] {$\hat{\mu}_{11}^{(4)}$} edge from parent node [above] {T}} child{node [dummy] {$\hat{\mu}_{21}^{(4)}$} edge from parent node [above] {F}};
		\end{tikzpicture} &
		\begin{tikzpicture}
  		[
   	 	grow                    = down,
    	sibling distance        = 8em,
    	level distance          = 4em,
    	edge from parent/.style = {draw, -latex},
    	every node/.style       = {font=\tiny}, sloped
  		]
  		\node [root] {$X_{i2}<-1.42$} child{node [env] {$X_{i1}<0.46$} child{node [dummy] {$\hat{\mu}_{12}^{(4)}$} edge from parent node [above] {T}} child{node [dummy] {$\hat{\mu}_{22}^{(4)}$} edge from parent node [above] {F}} edge from parent node [above] {T}} child{node [dummy] {$\hat{\mu}_{32}^{(4)}$} edge from parent node [above] {F}};
		\end{tikzpicture} &
		 \begin{tikzpicture}
  		[
   	 	grow                    = down,
    	sibling distance        = 8em,
    	level distance          = 4em,
    	edge from parent/.style = {draw, -latex},
    	every node/.style       = {font=\tiny}, sloped
  		]
  		\node [root] {$X_{i3}<0.48$} child {node [dummy] {$\hat{\mu}_{13}^{(4)}$}edge from parent node [above] {T}}
    		child {node [env] {$X_{i3}<1.56$} child {node [env] {$X_{i3}<0.76$} child{ node [env] {$X_{i1}<0.04$} child{node [dummy] {$\hat{\mu}_{23}^{(4)}$} edge from parent node [above] {T}} child{node [dummy] {$\hat{\mu}_{33}^{(4)}$} edge from parent node [above] {F}} edge from parent node [above] {T}} child{ node [dummy] {$\hat{\mu}_{43}^{(4)}$} edge from parent node [above] {F}} edge from parent node [above] {T}} child {node [dummy] {$\hat{\mu}_{53}^{(4)}$} edge from parent node [above] {F}} edge from parent node [above] {F}};
		\end{tikzpicture} &
		\begin{tikzpicture}
  		[
   	 	every node/.style       = {font=\tiny}, sloped
  		]
  		\node [dummy] {$\hat{\mu}_{14}^{(3)}$};
		\end{tikzpicture} \\
		\multicolumn{4}{c}{Iteration 5} \\
		Tree 1 & Tree 2 & Tree 3 & Tree 4 \\
		\begin{tikzpicture}
  		[
   	 	grow                    = down,
    	sibling distance        = 8em,
    	level distance          = 4em,
    	edge from parent/.style = {draw, -latex},
    	every node/.style       = {font=\tiny}, sloped
  		]
  		\node [root] {$X_{i2}<-0.58$} child{node [dummy] {$\hat{\mu}_{11}^{(5)}$} edge from parent node [above] {T}} child{node [env] {$X_{k2}<0.20$} child{node [dummy] {$\hat{\mu}_{21}^{(5)}$} edge from parent node [above] {T}} child{node [dummy] {$\hat{\mu}_{31}^{(5)}$} edge from parent node [above] {F}} edge from parent node [above] {F}};
		\end{tikzpicture} &
		\begin{tikzpicture}
  		[
   	 	grow                    = down,
    	sibling distance        = 8em,
    	level distance          = 4em,
    	edge from parent/.style = {draw, -latex},
    	every node/.style       = {font=\tiny}, sloped
  		]
  		\node [root] {$X_{i2}<-1.42$} child{node [env] {$X_{i1}<0.46$} child{node [dummy] {$\hat{\mu}_{12}^{(5)}$} edge from parent node [above] {T}} child{node [env] {$X_{i2}<-1.52$} child{node [dummy] {$\hat{\mu}_{22}^{(5)}$} edge from parent node [above] {F}} child{node [dummy] {$\hat{\mu}_{32}^{(5)}$} edge from parent node [above] {F}} edge from parent node [above] {F}} edge from parent node [above] {T}} child{node [dummy] {$\hat{\mu}_{42}^{(5)}$} edge from parent node [above] {F}};
		\end{tikzpicture} &
		 \begin{tikzpicture}
  		[
   	 	grow                    = down,
    	sibling distance        = 8em,
    	level distance          = 4em,
    	edge from parent/.style = {draw, -latex},
    	every node/.style       = {font=\tiny}, sloped
  		]
  		\node [root] {$X_{i3}<0.48$} child {node [dummy] {$\hat{\mu}_{13}^{(5)}$}edge from parent node [above] {T}}
    		child {node [env] {$X_{i3}<1.56$} child {node [env] {$X_{i3}<0.76$} child{node [dummy] {$\hat{\mu}_{23}^{(5)}$} edge from parent node [above] {T}} child{ node [dummy] {$\hat{\mu}_{33}^{(5)}$} edge from parent node [above] {F}} edge from parent node [above] {T}} child {node [dummy] {$\hat{\mu}_{43}^{(5)}$} edge from parent node [above] {F}} edge from parent node [above] {F}};
		\end{tikzpicture} &
		\begin{tikzpicture}
  		[
   	 	every node/.style       = {font=\tiny}, sloped
  		]
  		\node [dummy] {$\hat{\mu}_{14}^{(3)}$};
		\end{tikzpicture} \\
	\end{tabular}
\end{figure}
\end{landscape}

With the initializations in place, BART starts to draw the tree structures for each regression tree in the first MCMC iteration. Without loss of generality, let us start with determining $(T_1,M_1)$, the first regression tree. This is possible because the ordering of regression tree calculation does not matter. We first calculate $R_1={Y}-\sum_{j\neq 1}g(X,T_j,{M}_j)={Y}-[g(X,T_2,{M}_2)+g(X,T_3,{M}_3)+g(X,T_4,{M}_4)]={Y}-3\times\frac{\bar{{Y}}}{4}$. Then a MH algorithm is used to determine the posterior draw of the tree structure, $T_1$ for this iteration. The basic idea of MH is to propose a new tree structure from $T_1$, call this $T_1^*$, and then calculate the probability of whether $T_1^*$ should be accepted, taking into consideration: ${R}_1|T_1^*$ (the likelihood of the residual given the new tree structure), ${R}_1|T_1$ (the likelihood of the residual given the previous tree structure), the probability of observing $T_1^*$, the probability of observing $T_1$, the probability of moving from $T_1^*$ to $T_1$, and the probability of moving from $T_1$ to $T_1^*$. We describe the different types of moves from $T_1$ to $T_1^*$ in detail in the next subsection. If $T_1^*$ is accepted, $T_1$ is updated to become $T_1^*$ i.e. $T_1=T_1^*$. Else, nothing would be changed for $T_1$. From Figure \ref{bart_algo_mcmc13}, we can see that $T_1^*$ was not accepted in the first iteration so the tree structure remains as a single root node. The algorithm then updates ${M}_1$ based on the new updated regression structure for $T_1$ and moves on to determine $(T_2,{M}_2)$. 

To determine $(T_2,{M}_2)$, again the algorithm calculates ${R}_2={Y}-\sum_{j\neq 2}g(X,T_j,{M}_j)={Y}-[g(X,T_1,{M}_1)+g(X,T_3,{M}_3)+g(X,T_4,{M}_4)]={Y}-[\hat{\mu}_{11}^{(1)}+2\times\frac{\bar{{Y}}}{4}]$, where $\hat{\mu}_{11}^{(1)}$ is the updated parameter for regression tree 1. Similarly, MH is used to propose a new $T_2^*$ and ${R}_2$ is used to calculate the acceptance probability to decide whether $T_2^*$ should be accepted. Again, we see from Figure \ref{bart_algo_mcmc13} that $T_2^*$ was not accepted and hence a single parameter $\hat{\mu}_{12}^{(1)}$, drawn from ${M}_2|T_2,{R}_2,\sigma$, is used for $g(X,T_2,{M}_2)$. For $(T_3,{M}_3)$, the MH iteration result is more interesting because the newly proposed $T_3^*$ was accepted and we can see from Figure \ref{bart_algo_mcmc13} that a new tree structure was used for $T_3$ in Iteration 1. As a result, when calculating ${R}_4$, this becomes ${R}_4={Y}-[\hat{\mu}_{11}^{(1)}+\hat{\mu}_{21}^{(1)}+\hat{\mu}_{13}^{(1)}I\{{X}_3<0.48\}+\hat{\mu}_{23}^{(1)}I\{{X}_3\geq 0.48\}+\hat{\mu}_{14}^{(1)}]$. $T_4^*$ was not accepted and a single node $T_4$ was updated as the tree structure for $(T_4,{M}_4)$. Once the regression tree draws are complete, the BART then proceeds to draw the rest of the parameters in the BART model. More details in the next subsection.

Figures \ref{bart_algo_mcmc13} and \ref{bart_algo_mcmc45} give the full iterations from initiation to iteration 5. From these figures we can see how the four regression trees grow and change from one iteration of the MCMC to another. This iterative process runs for a burn-in period (typically 100 to 1000 iterations), before those draws are discarded, and then run for as long as needed to obtain a sufficient number of draws from the posterior distribution of $\sum_{j=1}^m g(X,T_j,{M}_j)$. After any full iteration in the MCMC algorithm, we have a full set of trees. We can therefore obtain a predicted value of $Y$ for any $X$ of interest (simply by summing the terminal node $\mu$'s). By obtaining predictions across many iterations, we also can easily obtain a 95\% prediction interval. Another point to note is how shallow the regression trees are in Figures \ref{bart_algo_mcmc13} and \ref{bart_algo_mcmc45} with a maximum depth of 3. This is because the regression trees are heavily penalized (via the prior) to reduce the likelihood for a single tree to grow very deep. This concept is borrowed from the idea that many weak models combined together performs much better than utilizing a very strong model which requires careful tweaking in order for the model to perform well.

\subsubsection{A rigorous perspective on the BART algorithm}
Now that we have a visual understanding of how the BART algorithm works, we shall give a more rigorous explanation of BART. First, we start with the prior distributions for BART. The prior distribution for Equation (\ref{bart_con}) is $P(T_1,{M}_1,\ldots,T_m,{M}_m,\sigma)$. The usual prior specification is that $\{(T_1,{M}_1)$, $\ldots$, $(T_m,{M}_m)\}$ and $\sigma$ are independent and that $(T_1,{M}_1),\ldots,(T_m,{M}_m)$ are independent of each other. Then the prior distribution can be written as
\begin{align}
	\label{prior_decom}
	P(T_1,{M}_1,\ldots,T_m,{M}_m,\sigma)&=P(T_1,{M}_1,\ldots,T_m,{M}_m)P(\sigma)\nonumber\\
	&=[\prod_j^mP(T_j,{M}_j)]P(\sigma)\nonumber\\
	&=[\prod_j^mP({M}_j|T_j)P(T_j)]P(\sigma)\nonumber\\
	&=[\prod_j^m\{\prod_k^{b_j}P(\mu_{kj}|T_j)\}P(T_j)]P(\sigma).
\end{align}
For the third to fourth line in Equation (\ref{prior_decom}), recall that ${M}_j=\{\mu_{1j},\ldots,\mu_{b_jj}\}$ is the vector of terminal node parameters associated with $T_j$ and each node parameter $\mu_{kj}$ is usually assumed to be independent of each other. Equation (\ref{prior_decom}) implies that we need to set distributions for the priors $\mu_{kj}|T_j$, $\sigma$, and $T_j$. The priors for $\mu_{kj}|T_j$ and $\sigma$ are usually given as
$\mu_{kj}|T_j\sim N(\mu_{\mu},\sigma_{\mu}^2)$ and $\sigma^2\sim IG(\frac{\nu}{2},\frac{\nu\lambda}{2})$ respectively, where $IG(\alpha,\beta)$ is the inverse gamma distribution with shape parameter $\alpha$ and rate parameter $\beta$.

The prior for $P(T_j)$ is more interesting and can be specified using three aspects:
\begin{enumerate}
	\item The probability that a node at depth $d=0,1,\ldots$ would split, which is given by $\frac{\alpha}{(1+d)^{\beta}}$. The parameter $\alpha\in\{0,1\}$ controls how likely a node would split, with larger values increasing the likelihood of a split. The number of terminal nodes is controlled by parameter $\beta>0$, with larger values of $\beta$ reducing the number of terminal nodes. This aspect is important as this is the penalizing feature of BART which prevents BART from overfitting and allowing convergence of BART to the target function $f(X)$ \citep{rockova_bart}. As mentioned in the previous subsection, this aspect also allows many shallow (weak) regression trees to be fit and eventually summed together to obtain a stronger model.
	\item The distribution used to select the covariate to split upon in an internal node. The default suggested distribution is the uniform distribution. Recent work \citep{rockova_bcart,linero} have argued that the uniform distribution does not promote variable selection and should be replaced if variable selection is desired.
	 \item The distribution used to select the cutoff point in an internal node once the covariate is selected. The default suggested distribution is the uniform distribution.  
\end{enumerate}
The setting of the hyper-parameters for the BART priors is rather technical so we refer interested readers to our Appendix for how this can be done. 

The prior distribution would induce the posterior distribution
\begin{align*}
	P[(T_1,{M}_1),\ldots,(T_m,{M}_m),\sigma|{Y}]&\propto P({Y}|(T_1,{M}_1),\ldots,(T_m,{M}_m),\sigma)\\&\quad\times P((T_1,{M}_1),\ldots,(T_m,{M}_m),\sigma)
\end{align*}
which can be simplified into two major posterior draws using Gibbs sampling. First, draw $m$ successive 
\begin{equation}
	\label{tree_draws}
	P[(T_j,{M}_j)|T_{(j)},{M}_{(j)},{Y},\sigma]
\end{equation}
for $j=1,\ldots,m$, where $T_{(j)}$ and ${M}_{(j)}$ consist of all the tree structures and terminal nodes except for the $j^{\text{th}}$ tree structure and terminal node; then, draw 
\begin{equation}
	\label{sigma_draws}
	P[\sigma|(T_1,{M}_1),\ldots,(T_m,{M}_m),{Y}]
\end{equation}
from $IG(\frac{\nu+n}{2},\frac{\nu\lambda+\sum_{i=1}^n(Y_i-\sum_{j=1}^m g(X_i,T_j,{M}_j))^2}{2})$.

To obtain a draw from (\ref{tree_draws}), note that this distribution depends on $(T_{(j)},{M}_{(j)},{Y},\sigma)$ through
\begin{equation}
	{R}_j={Y}-\sum_{w\neq j}g(X,T_w,{M}_w),
\end{equation}
the residuals of the $m-1$ regression sum of trees fit excluding the $j^{\text{th}}$ tree (Recall our visual example in the previous subsection). Thus (\ref{tree_draws}) is equivalent to the posterior draw from a single regression tree $R_{ij}=g({X}_i,T_j,{M}_j)+\varepsilon_i$ or 
\begin{equation}
	\label{tree_draws_red}
	P[(T_j,{M}_j)|{R}_j,\sigma].
\end{equation}
We can obtain a draw from (\ref{tree_draws_red}) by first integrating out ${M}_j$ to obtain $P(T_j|{R}_j,\sigma)$. This is possible since a conjugate Normal prior on $\mu_{kj}$ was employed. We draw $P(T_j|{R}_j,\sigma)$ using a MH algorithm where first, we generate a candidate tree $T_j^*$ for the $j^{\text{th}}$ tree with probability distribution $q(T_j,T_j^*)$ and then, we accept $T_j^*$ with probability 
\begin{equation}
	\label{bcart_mh}
	\alpha(T_j,T_j^*)=\min\left\{1,\frac{q(T_j^*,T_j)}{q(T_j,T_j^*)}\frac{P({R}_j|X,T_j^*,M_j)}{P({R}_j|X,T_j,M_j)}\frac{P(T_j^*)}{P(T_j)}\right\}.
\end{equation}
$\frac{q(T_j^*,T_j)}{q(T_j,T_j^*)}$ is the ratio of the probability of how the previous tree moves to the new tree against the probability of how the new tree moves to the previous tree, $\frac{P({R}_j|X,T_j^*,M_j)}{P({R}_j|X,T_j,M_j)}$ is the likelihood ratio of the new tree against the previous tree, and $\frac{P(T_j^*)}{P(T_j)}$ is the ratio of the probability of the new tree against the previous tree.

A new tree $T_j^*$ can be proposed given the previous tree $T_j$ using four local steps: (i) grow, where a terminal node is split into two new child nodes; (ii) prune, where two terminal child nodes immediately under the same non-terminal node are combined together such that their parent non-terminal node becomes a terminal node; (iii) swap, the splitting criteria of two non-terminal nodes are swapped; (iv) change, the splitting criteria of a single non-terminal node is changed. Once we have the draw of $P(T_j|{R}_j,\sigma)$, we then draw $P(\mu_{kj}|T_j,{R}_j,\sigma)\sim N(\frac{\sigma_{\mu}^2\sum_{k=1}^{n_k}R_{kj}}{n_k\sigma_{\mu}^2+\sigma^2},\frac{\sigma^2\sigma_{\mu}^2}{n_k\sigma_{\mu}^2+\sigma^2})$, where $R_{kj}$ is the subset of elements in ${R}_j$ allocated to the terminal node parameter $\mu_{kj}$ and $n_k$ is the number of $R_{kj}$s allocated to $\mu_{kj}$. We derive $P(\mu_{kj}|T_j,{R}_j,\sigma)$, Equation (\ref{sigma_draws}), and Equation (\ref{bcart_mh}) for the grow and prune steps as an example in our Appendix.

\subsection{Binary outcomes}
For binary outcomes, BART can be extended using a probit model. Specifically,
\begin{equation*}
	P[Y_i=1|X_i,(T_1,{M}_1),\ldots,(T_m,{M}_m)]=\Phi[\sum_{j=1}^m g(X_i;T_j,{M}_j)]
\end{equation*} 
where $\Phi[.]$ is the cumulative distribution function of a standard normal distribution and $i$ indexes the subjects $i=1,\ldots,n$. With such a setup, only priors for $(T_1,{M}_1),\ldots,(T_m,{M}_m)$ are needed. The same decomposition in Equation (\ref{prior_decom}) without $\sigma$ can be employed and the similar prior specifications for $\mu_{kj}|T_j$ and $T_j$ can be used. The setup of the hyper-parameters are slightly different from that of continuous outcomes and we describe this in the Appendix.

To estimate the posterior distribution, data augmentation \citep{albert_chib_1993} can be used. In essence, we first draw a latent variable ${Z}=\{Z_1,\ldots,Z_n\}$ as follows:
\begin{align*}
	Z_i\sim N_{(-\infty,0)}[\sum_{j=1}^m g(X_i;T_j,{M}_j),1]&\quad\text{if }Y_i=0,\\
	Z_i\sim N_{(0,\infty)}[\sum_{j=1}^m g(X_i;T_j,{M}_j),1]&\quad\text{if }Y_i=1
\end{align*}
where $N_{(a,b)}[\mu,\sigma^2]$ is a truncated normal distribution with mean $\mu$ and variance $\sigma^2$ truncated at $(a,b)$. Next, we can treat ${Z}$ as the continuous outcome for a BART model with
\begin{equation}
	\label{bin_bart_mod}
	{Z}=\sum_{j=1}^m g(X;T_j,{M}_j)+\varepsilon
\end{equation}
where $\varepsilon\sim N(0,1)$ because we employed a probit link. The usual posterior estimation for a continuous outcome BART with $\sigma\equiv 1$ can now be employed on Equation (\ref{bin_bart_mod}) for one iteration in the MCMC. The updated $\sum_{j=1}^m g(X;T_j,{M}_j)$ can then be used to draw a new ${Z}$ and this new ${Z}$ can be used to draw another iteration of $\sum_{j=1}^m g(X;T_j,{M}_j)$. The process can then be repeated till convergence.

\section{Illustrating the performance for BART}
\label{bart_app}
\subsection{Posterior performance via synthetic data}
\begin{landscape}
\begin{figure}
\caption{Posterior mean and 95\% credible interval of Bayesian linear regression (BLR) and BART with $m=1,50,100,150,200$ for 20 randomly selected testing set outcomes. $n=1,000$, Black=true value, colored=model estimates. \label{sim_res}}
	\begin{tabular}{ccc}
	BLR & BART with $m=1$ & BART with $m=50$\\
	\includegraphics[scale=0.3]{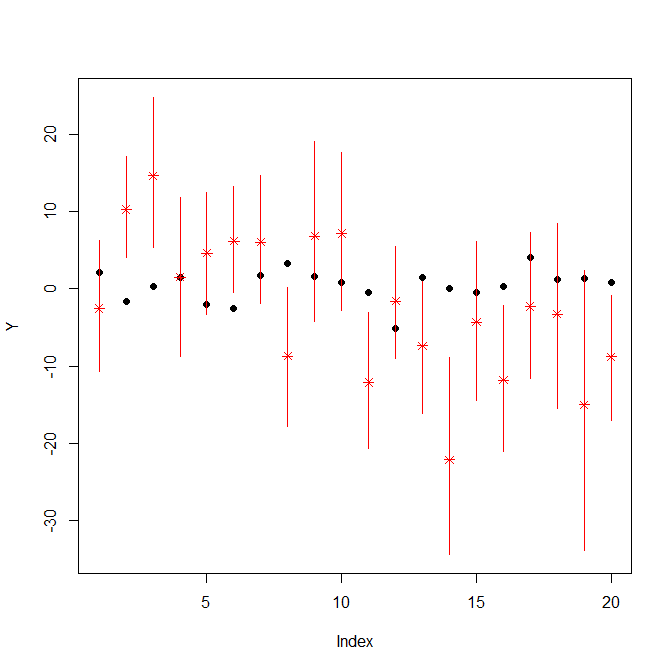} & \includegraphics[scale=0.3]{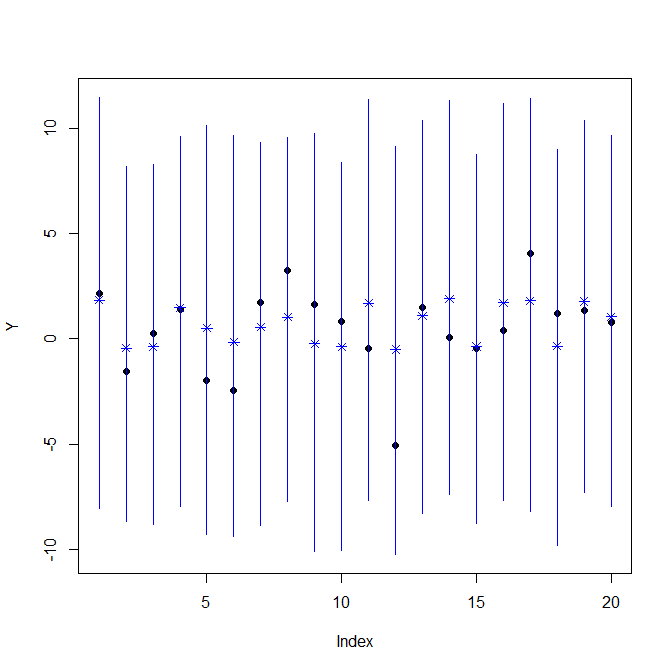} & \includegraphics[scale=0.3]{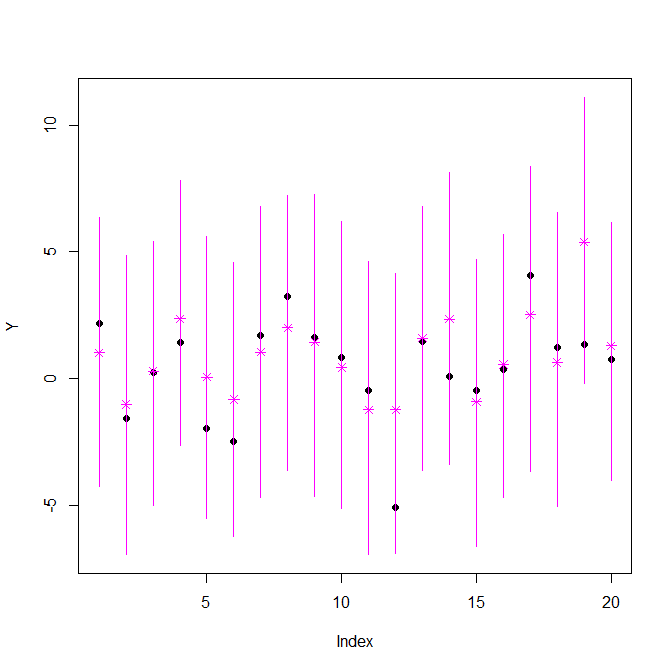} \\
	BART with $m=100$ & BART with $m=150$ & BART with $m=200$\\
	\includegraphics[scale=0.3]{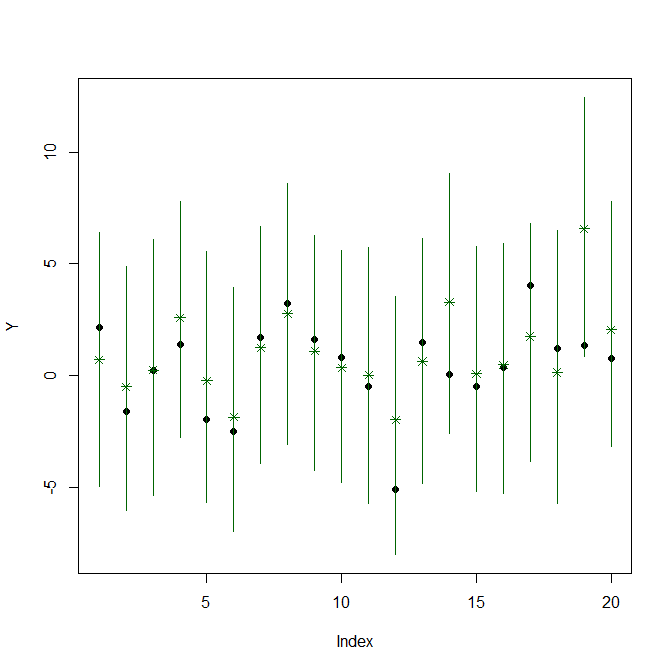} & \includegraphics[scale=0.3]{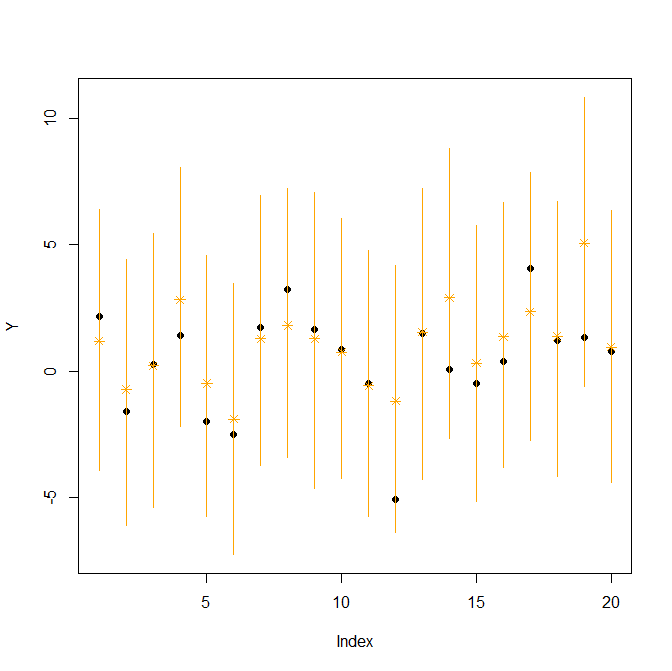} & \includegraphics[scale=0.3]{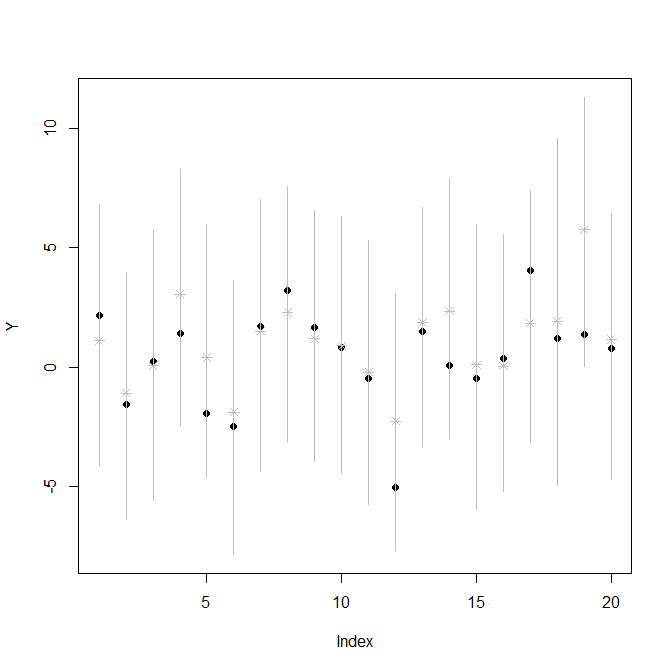} \\
	\end{tabular}
\end{figure}
\end{landscape}

We generated a synthetic data set with $p=3$, $n=1,000$ and, the true model for $Y_i$ is
\begin{align*}
	Y_i&=0.5+0.1X_{i1}+0.3X_{i2}^2+0.7\sin(X_{i3})+0.2X_{i1}X_{i2}+0.9\sqrt{|X_{i1}X_{i3}|}\nonumber\\
	&\quad+0.4\exp(X_{i2}X_{i3})+0.8\log(|X_{i1}X_{i2}X_{i3}|)+\varepsilon_i
\end{align*} 
with $X_{ip}\sim N(0,1)$ and $\varepsilon_i\sim N(0,2)$. The goal is to demonstrate that BART can predict $Y$'s effectively even in complex, non-linear models, and also properly accounts for prediction uncertainty compared to a parametric BLR model. To this end, we randomly selected 880 samples as the training set and then use the remaining 20 samples as the testing set. We also varied the number of trees used by BART to illustrate how varying $m$ affects the performance of BART. We plotted the point estimate and 95\% credible interval of the 20 randomly selected testing data points and compared them with their true values in Figure \ref{sim_res}. The codes to implement this simulation will be made available on https://github.com/yaoyuanvincent.

We can see from Figure \ref{sim_res} that most of the point estimates of BLR were far away from their true values and many of the true values were not covered by the 95\% credible interval. For BART with a single tree, although the true values were mostly covered by the 95\% credible interval, the point estimates were far from their true values. When we increased the number of trees to 50 in BART, we see a significant improvement in terms of bias (closeness to the true values) compared to both BLR and BART with $m=1$. In addition, we see a narrowing of the 95\% intervals. We see that as we increase the number of trees, the point estimate and 95\% intervals stabilize. In other words, we might see a big difference between $m=1$ and $m=50$, and virtually no difference between $m=200$ and $m=20,000$. In practice, the idea is to choose a large enough value for $m$ (default is often 200) so that it very well approximates the results that would have been obtained if more trees were used. One way to determine an $m$ that is sufficiently large is with cross validation \citep{chipman_bart}.
 
\subsection{Predicting the Standardized Hospitalization Ratio from the 2013 Centers for Medicare and Medicaid Services Dialysis Facility Compare dataset}
We next present an example to demonstrate how BART can be applied to a dataset to improve prediction over the usual multiple linear regression model. The 2013 Centers for Medicare and Medicaid Services Dialysis Facility Compare dataset contains information regarding 105 quality measures and 21 facility characteristics of all dialysis facilities in the US, including US territories. This dataset is available publicly at https://data.medicare.gov/data/archives/dialysis-facility-compare. The codes and data to implement this analysis will be made available on https://github.com/yaoyuanvincent. We are interested in finding a model that can better predict the standardized hospitalization ratio (SHR). This quantity is important because a large portion of dialysis cost for End Stage Renal Disease (ESRD) patients can be attributed to patient hospitalizations. 

\begin{table}[H]
\caption{Descriptive statistics of dialysis facility characteristics and quality measures (n=5,774). \label{dstats}}
\begin{center}
	\hspace*{-2cm}
	\footnotesize
	\begin{tabular}{|lr|lr|}
	\hline 
Parameters (\% missing) & Mean (s.d)/Frequency (\%) & Parameters (\% missing) & Mean (s.d)/Frequency (\%) \\
\hline
Arterial Venous Fistula (3) & 63.27 (11.22) & Number of stations & 18.18 (8.27)\\
Avg. Hemoglobin$<$10.0 g/dL (5) & 12.86 (10.32) & Serum P. 3.5-4.5mg/dL$^{**}$ (2) & 28.52 (5.13) \\
Chain name: & & Shift after 5pm? & \\
\hspace*{0.5cm} Davita & 1,812 (31) & \hspace*{0.5cm} Yes & 1,097 (19)\\
\hspace*{0.5cm} FMC & 1,760 (30) & \hspace*{0.5cm} No & 4,677 (81)\\
\hspace*{0.5cm} Independent & 820 (14) & SHR & 1.00 (0.31)\\
\hspace*{0.5cm} Medium & 740 (13) & SMR (2) & 1.02 (0.29)\\
\hspace*{0.5cm} Small & 642 (12) & STR (7) & 1.01 (0.54)\\
Patient volume$^*$ & 100.07 (60.93) & Type: & \\
Facility Age (years) & 14.47 (9.81) & \hspace*{0.5cm} All (HD, Home HD, \& PD) & 1,443 (25) \\
For profit? & & \hspace*{0.5cm} HD \& PD & 1,897 (33)\\
\hspace*{0.5cm} Yes & 4,967 (86) & \hspace*{0.5cm} HD \& Home HD & 103 (2)\\
\hspace*{0.5cm} No & 806 (14) & \hspace*{0.5cm} HD alone & 2,331 (40)\\
HD$\geq$1.2 Kt/V (4) & 88.52 (9.85) & URR$\geq$65\% (7) & 98.77 (3.04)\\
Hypercalcemia (3) & 2.37 (3.20) & Vas. Catheter$>$90 days (3) & 10.74 (6.66)\\
\hline
	\end{tabular} 
\end{center}
\footnotesize *Estimated. **Normal range.
\end{table}

Table \ref{dstats} shows some descriptive statistics for this dataset. SHR was adjusted for a patient’s age, sex, duration of ESRD, comorbidities, and body mass index at ESRD incidence. We removed 463 facilities (7\%) with missing SHR values because of small patient number. We also removed peritoneal dialysis (PD) removal greater than 1.7 Kt/V because of the high proportion of missingness (80\%). We combined pediatric hemodialysis (HD) removal greater than 1.2 Kt/V with adult HD removal greater than 1.2 Kt/V because most facilities (92\%) do not provide pediatric HD. We re-categorized the chain names to ``Davita,'' ``Fresenius Medical Care (FMC),'' ``Independent,'' ``Medium,'' and ``Small.'' ``Medium'' consists of chains with 100-500 facilities while ``Small'' are chains with less than 100 facilities. To estimate patient volume, we used the maximum of the number of patients reported by each quality measure group: Urea Reduction Ratio (URR), HD, PD, Hemoglobin (HGB), Vascular Access, SHR, SMR, STR, Hypercalcemia (HCAL), and Serum phosphorus (SP). We also logarithm-transformed (log) SHR, SMR, and STR so that the theoretical range for these log standardized measures will be $-\infty$ to $\infty$. 

\begin{figure}[H]
	\caption{Root mean squared error for the 10-fold cross-validation of multiple linear regression (MLR), random forest, and Bayesian additive regression trees of log transformed standardized hospitalization ratio (SHR) \label{rmse_eg}}
	\centering
	\includegraphics[scale=0.75]{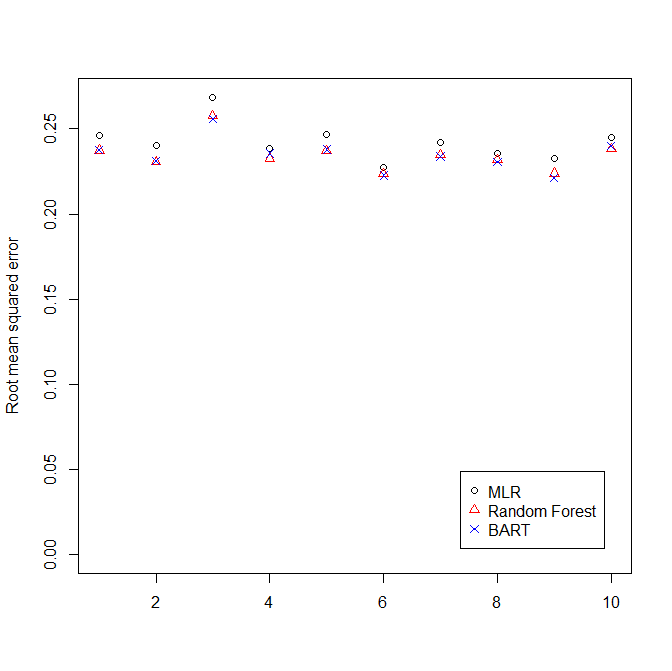}
\end{figure}

For our analysis, we used the log-transformed SHR as the outcome and the variables in Table \ref{dstats} as the predictors. We used the root mean squared error (RMSE) of a 10-fold cross-validation to compare the prediction performance from multiple linear regression (MLR), Random Forest (RF), and BART. For RF and BART, we used the default settings from the $R$ packages \textit{randomForest} and \textit{BayesTree} respectively. The 10 RMSEs produced by each method from the 10-fold cross validation is provided in Figure \ref{rmse_eg}. It is clear from this figure that BART and RF produce very similar prediction performances and is better compared to MLR. The mean of these 10 values also suggested a similar picture with MLR producing a mean of 0.24 while RF and BART produced a mean of 0.23.

\subsection{Predicting left turn stops at an intersection}
We next present another example where BART showed improvement in the prediction performance of a binary outcome. In \cite{tan_aap}, the authors were interested in predicting whether a human driven vehicle would stop at an intersection before making a left turn. Left turns are important in countries with right side driving because most vehicle conflicts including crashes at intersections occur during left turns. Knowledge of whether a human driven vehicle would stop before executing a left turn would help driverless vehicles be better able to make decisions at an intersection. More details about this dataset can be found in \cite{tan_aap}. In brief, the data comes from the Integrated Vehicle Based Safety System (IVBSS) study conducted by \cite{sayer}. This study collected driving data from 108 licensed drivers in Michigan between April 2009 and April 2010. Each driver drove one of the sixteen research vehicles fitted with various recording devices to capture the vehicle dynamics while the subject is driving on public roads for 12 days. In particular, \cite{tan_aap} focused on the vehicle speeds of all left turns at an intersection starting from 100 meters away from the center of an intersection to the center of an intersection. They then transformed the vehicle speed time series to a distance series. Having the vehicle speed at each distance as the columns and each turn as the rows, they then performed principal components analysis (PCA) on these vehicle speeds using moving windows of 6 meters from 94 meters away to 1 meter away from the center of an intersection. This implies that at each meter, a PCA analysis was conducted using 6 meters of vehicle speeds, i.e. at 94 meters, 94 to 100 meters away was used, at 93 meters, 93 to 99 meters away was used and so on until 1 meter away. They used a 6 meter moving window because they found that longer windows did not improve prediction performance and a 6 meter moving window provided the best prediction performance. At each meter, the first three principal components (PCs) from the corresponding 6 meter moving window PCAs were then used to determine the prediction model with the outcome as whether the vehicle stopped (vehicle speed $<1m/s$) in the future with stopped coded as 1 and not stop coded as 0. Only the first three PCs were used because these three PCs explained nearly 99\% of the variance in the 6 meter moving window distance series of vehicle speeds as well as provided the best prediction performance. This setup resulted in 94 models corresponding with 94 datasets for each meter. In order to keep our presentation concise, we focus on the dataset halfway through the turn maneuver (50 meters away from the center of an intersection) which is made up of the first 3 PCs of the PCA on vehicle speed from 50 to 56 meters and the outcome of whether the vehicle stopped in the future from 49 meters to the center of an intersection. We ran a 10-fold cross-validation on this dataset and compared the binary prediction results of logistic regression, RF, and BART. Since the outcome of interest for this dataset was binary, we used the area under the receiver operating curve (AUC) to determine the prediction performance instead of the RMSE, which is more suited for continuous outcomes.

\begin{figure}[H]
	\caption{Area under the receiver operating characteristic curve for the 10-fold cross-validation of logistic regression, random forest, and Bayesian additive regression trees of left turn stop probabilities at an intersection. \label{aap_auc}}
	\includegraphics[scale=0.75]{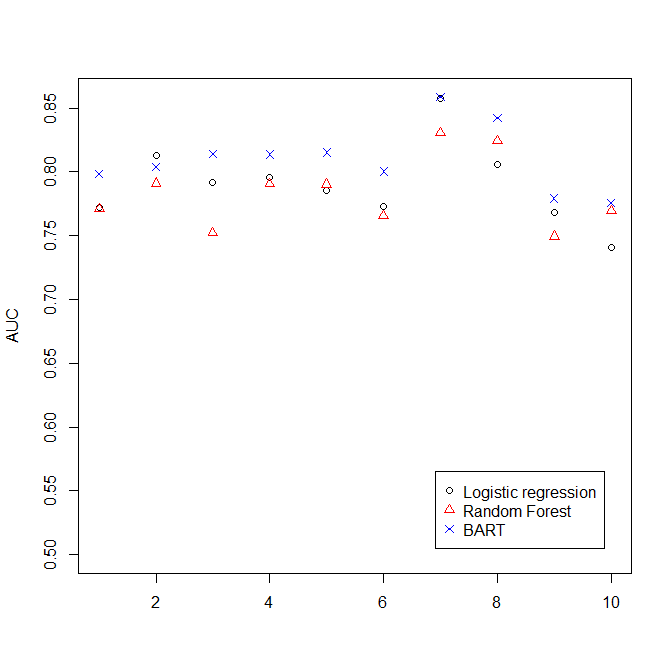}
\end{figure}

Figure \ref{aap_auc} shows the results of the 10 AUCs produced by each method from the 10-fold cross validation. It is clear that BART performed extremely well in predicting whether the human-driven vehicle would stop in the future at an intersection before making a left turn, much better than either logistic regression or RF. This is also evident from the mean of the 10 cross validation AUC values produced by each method. BART produced a mean of 0.81 compared to 0.79 from logistic regression and 0.78 from RF.

\section{General BART model}
\label{gen_bart}
Recently researchers have extended or generalized BART to a wider variety of settings, including clustered data, spatial data, semi-parametric models, and to situations where more flexible distributions for the error term is needed. Here we describe a more general BART framework that includes all of these cases and more. An important feature of this general BART model is they can be fitted without very extensive re-derivation for the MCMC draws of the regression trees described in Section \ref{review}. That is, the MCMC algorithm we described previously only needs small adjustments to handle this more general setting.

To set up our General BART model, suppose once again that we have a continuous outcome ${Y}$ and $p$ covariates $X=\{{X}_1,\ldots,{X}_p\}$. Suppose also that we have another set of $q$ covariates $W=\{{W}_1,\ldots,{W}_q\}$ such that no two columns in $X$ and $W$ are the same. Then, we can extend Equation (\ref{bart_con}) as follows:
\begin{equation}
	\label{joint_bart}
	{Y}=G[X,(T,M)]+H(W,{\Theta})+\varepsilon
\end{equation}
where $G[X,(T,M)]=\sum_{j=1}^m g(X;T_j,{M}_j)$ with $(T,M)=[(T_1,{M}_1),\ldots,(T_m,{M}_m)]$, $H(.)$ is a function that works on $W$ using parameter $\Theta$, and $\varepsilon\sim G(\Sigma)$ can be any distribution with parameter $\Sigma$. 

Assuming that $(T,M)$, ${\Theta}$, and ${\Sigma}$ are independent, the prior distribution for Equation (\ref{joint_bart}) is $P(T,M)P({\Theta})P({\Sigma})$. Assuming again that the $(T_j,{M}_j)$'s within $(T,M)$ are independent of each other, $P(T,M)$ can be decomposed into $\prod_j^m\{\prod_{k=1}^{b_j}P(\mu_{kj}|T_j)\}P(T_j)$. The priors needed are thus $P(\mu_{kj}|T_j)$, $P(T_j)$, $P({\Theta})$, and $P({\Sigma})$. Note that it is possible to model ${\Theta}$ and ${\Sigma}$ jointly so that the prior distribution becomes $\prod_j^m\{\prod_{k=1}^{b_j}P(\mu_{kj}|T_j)\}P(T_j)P({\Theta},{\Sigma})$ instead. We shall see this in Example \ref{dpmbart}.

To obtain the posterior distribution of $P[(T,M),{\Theta},{\Sigma}|{Y},X,W]$, Gibbs sampling can be used. For $P[(T,M)|{\Theta},{\Sigma},{Y},X,W]$, this can be seen as drawing from the following model
\begin{equation}
	\label{G_draw}
	{\tilde{Y}}=G[X,(T,M)]+\varepsilon
\end{equation}
where ${\tilde{Y}}={Y}-H(W,{\Theta})$ which is just a BART model with a modified outcome ${\tilde{Y}}$. Hence, the BART algorithm presented in Section \ref{review} can be used to draw $(T,M)$, the regression trees. Similarly, $P[{\Theta}|(T,M),{\Sigma},{Y},X,W]$ can be obtained by drawing from the model 
\begin{equation}
	\label{H_draw}
	{Y'}=H(W,{\Theta})+\varepsilon
\end{equation}
where ${Y'}={Y}-G[X,(T,M)]$. This posterior draw depends on the function $H(.)$ being used as well as the prior distribution specified for ${\Theta}$. As there are many possibilities where we can set up $H(.)$ and ${\Theta}$, we shall not discuss the specifics here. The examples we present in the subsequent subsections will highlight a few of these possibilities we have seen in the literature thus far. Finally, drawing from $P[{\Sigma}|(T,M),{\Theta},{Y},X,W]$ is just drawing from the model 
\begin{equation}
	\label{e_draw}
	{R}=\varepsilon
\end{equation}
where ${R}={Y}-G[X,(T,M)]-H(W,{\Theta})$. Again, many possibilities are available for setting up the prior distribution for $\Sigma$ and hence the distributional assumption for $\varepsilon$. The default is usually $e_i\sim N(0,\sigma^2)$ where ${\Sigma}=\sigma^2\sim IG(\frac{\nu}{2},\frac{\nu\lambda}{2})$. Example \ref{dpmbart} shows an alternative distributional assumption for $\varepsilon$ and hence, $\Sigma$. Iterating through the above three Gibbs steps will give us the posterior draw of $P[(T,M),{\Theta},{\Sigma}|{Y},X,W]$. 

For binary outcomes, the probit link can once again be used where 
\begin{equation*}
	P[Y_i=1|X,(T_1,{M}_1),\ldots,(T_m,{M}_m)]=\Phi[G\{X,(T,M)\}+H(W,{\Theta})].
\end{equation*} 
Under this framework, we will only need priors for $P(T,M)$ and $P({\Theta})$. $P(T,M)$ can be decomposed once again into $\prod_j^m\{\prod_{k=1}^{b_j}P(\mu_{kj}|T_j)\}P(T_j)$ if we are willing to assume that the $m$ trees are independent of one another, and data augmentation \citep{albert_chib_1996} can be used obtain the posterior distribution. We can draw
\begin{align*}
	Z_i\sim N_{(-\infty,0)}[G[X,(T,M)]+H(W,{\Theta}),1]&\quad\text{if }Y_i=0,\\
	Z_i\sim N_{(0,\infty)}[G[X,(T,M)]+H(W,{\Theta}),1]&\quad\text{if }Y_i=1
\end{align*}
and then treat ${Z}$ as the outcome for the model in Equation (\ref{joint_bart}). This would imply that $\varepsilon_k\sim N(0,1)$ in Equation (\ref{joint_bart}) and we can apply the Gibbs sampling procedure we described for continuous outcomes using ${Z}$ instead of ${Y}$ with ${\Sigma}=\sigma=1$. Iterating through the latent draws and Gibbs steps will produce the posterior distribution that we require.

With the general framework and model for BART in place, we are now equipped to consider how \cite{roy_semibart}, \cite{tan_ribart}, \citep{zhang}, and \cite{george} extended BART to solve their research problems in the next fours subsections.

\subsection{Semiparametric BART} 
\label{semibart}
The semiparametric BART was first presented by \cite{roy_semibart}. Their idea was to have a model where the effects of interest are modeled linearly with at most simple interactions to keep the associated parameters interpretable while having the nuisance or confounder variables be modeled as flexibly as possible. In its simplest form, we have under the framework of Equation (\ref{joint_bart}) that
\begin{equation*}
	H(W,{\Theta})=\theta_0+\theta_1{W}_1+\ldots+\theta_q{W}_q
\end{equation*} 
where $W=\{{W}_1,\ldots,{W}_q\}$, ${\Theta}=\{\theta_0,\ldots,\theta_q\}$, and $\varepsilon_i\sim N(0,\sigma^2)$ with $\Sigma=\sigma$. Prior distributions for $\mu_{kj}|T_j$, $T_j$, and $\sigma^2$ follow the usual distributions we use for BART while $\Theta\sim MVN(\beta,\Omega)$, possibly. Posterior estimation follows the procedure we described in Section \ref{gen_bart} using Gibbs Sampling. For Equations (\ref{G_draw}) and (\ref{e_draw}), since they suggested using the default BART priors, the usual BART mechanisms can be applied to obtain the posterior draws. For Equation (\ref{H_draw}), ${\Theta}\sim MVN(\beta,\Omega)$ implies that we can treat this as the usual BLR and standard Bayesian methods could be used to obtain the posterior draw for ${\Theta}$. The framework for binary outcomes follows easily using the data augmentation step we describe in Section \ref{gen_bart}.   

\subsection{Random intercept BART for correlated outcomes} 
\label{ribart}
Random intercept BART (riBART) was proposed by \cite{tan_ribart} as a method to handle correlated continuous or binary outcomes with correlated binary outcomes as the main focus. Under the framework of Equation (\ref{joint_bart}), we have $H(W,{\Theta})=W{a}$, where ${\Theta}=({a},\tau)$ and
\begin{equation*}
	W=\left[\begin{matrix}
		1 & 0 & \ldots & 0 \\
		\vdots & \vdots & \ddots & \vdots \\
		1 & 0 & \ldots & 0 \\
		0 & 1 & \ldots & 0 \\
		\vdots & \vdots & \ddots & \vdots \\
		0 & 1 & \ldots & 0 \\
		\vdots & \vdots & \vdots & \vdots \\
		0 & 0 & \ldots & 1 \\
		\vdots & \vdots & \ddots & \vdots \\
		0 & 0 & \ldots & 1 \\
	\end{matrix}\right]
\end{equation*}
i.e. $W$ is a matrix with 1 repeated $n_1$ times in the first column and 0s for the rest of the column, 0 repeated $n_1$ times in the second column followed by 1 repeated $n_2$ times and then 0s for the rest of the columns, and so on until for the last column we have 0 repeated $\sum_{l=1}^{L-1}n_l$ times and then 1 repeated $n_L$ times, ${a}=\{a_1,\ldots,a_L\}$ with $a_l|\tau^2\sim N(0,\tau^2)$. $l$ indexes the subject. Once again, $\varepsilon\sim N(0,\sigma^2)$ with $\Sigma=\sigma$ and the usual BART priors for $\sigma$, $\mu_{kj}|T_j$, and $T_j$ can be employed. $a_l$ and $\varepsilon$ are assumed to be independent. A simple prior of $\tau^2\sim IG(1,1)$ could be used although more robust or complicated priors are possible. Posterior estimation and binary outcomes then follow the procedure described in Section \ref{gen_bart} easily.

\subsection{Spatial BART for a statistical matched problem} 
\label{spatialbart}
The Spatial BART approach of \cite{zhang} was proposed to handle statistical matched problems \citep{rassler} that occur in surveys. In statistical matched problems, inference is desired for the relationship between two different variables collected by two different datasets on the same subject. For example, survey A may collect information on income but survey B collects information on blood pressure. Both surveys A and B contain subjects that overlap. The relationship between income and blood pressure is then desired. To solve this problem, Spatial BART essentially uses a framework similar to that of riBART described in Example \ref{ribart} with a more complicated prior distribution for ${\Theta}$. The specification for $W$ and ${a}$ is the same but the distribution placed on ${a}$ is instead the conditionally autoregressive prior which can be specified as
\begin{equation}
	\label{sbart}
	{a}|\rho,\delta^2\sim N(0,\delta^2({H}-\rho{C})^{-1})
\end{equation}
where ${C}={c_{il}}$ is a $I\times I$ adjacency matrix, $l=1,\ldots,I$, with $c_{il}=1$ if group $i$ and group $l$ are (spatial) neighbors for $i\neq l$; $c_{il}=0$ otherwise; and $c_{il}=0$ if $i=l$. $H$ is a diagonal $I\times I$ matrix with diagonals $h_i=\sum_{l=1}^I c_{il}$, $\rho$ is a parameter with range $(-1,1)$, and $\delta^2$ is the variance component for Equation (\ref{sbart}). $\rho$ and $\delta^2$ are hyperparameters that is prespecified. Finally, $\varepsilon\sim N(0,\sigma^2)$ and Equation (\ref{joint_bart}) is completed by placing the usual BART priors for $\sigma$, $\mu_{kj}|T_j$, and $T_j$. Posterior draws again follow the procedures we outlines in Section \ref{gen_bart}.

\subsection{Dirichlet Process Mixture BART} 
\label{dpmbart}
The Dirichlet Process Mixture (DPM) BART was proposed by \cite{george} to enhance the robustness of distributional assumption for $\varepsilon$ in Equation (\ref{bart_con}). To do this they focused on a different specification for $\varepsilon$ by assuming that
\begin{align*}
	\varepsilon_i&\sim N(a_i,\sigma_i^2), \\
	(a_i,\sigma_i^2)&\sim D,\\
	D&\sim DP(D_0,\alpha)
\end{align*}
where $D$ denote a random discrete distribution and $DP$ denotes the Dirichlet process with parameters $D_0$ and $\alpha>0$. The atoms of $D$ can be seen as iid draws from $D_0$. $\alpha$ on the other hand determines weight allocated to atom of discrete $D$. Higher values of $\alpha$ imply that the weights would be spread out among the atoms. Lower values of $\alpha$ imply that weights would be concentrated on only a few atoms. Although the assumption of $\varepsilon_i\sim N(a_i,\sigma_i^2)$ suggests that each subject will have their own mean and variance for the error term, the placement of a Dirichlet process on $D$ restricts the number of unique components for $(a_i,\sigma_i^2)$ to $K<n$, which ensures that this model would still be identifiable. Viewing DPMBART as a form of Equation (\ref{joint_bart}), we have $H(W,{\Theta})=Wa$ where $W$ and $a$ have the same structure as riBART and $P({\Theta},{\Sigma})=(a_i,\sigma_i^2)$. Note here that we are no longer assuming that $a_i$ and $\varepsilon_i$ are independent unlike in some of our previous examples.

The priors for DPMBART are $D_0$ and $\alpha$. For $D_0$, the commonly employed form is $P(\mu,\sigma|\nu,\lambda,\mu_0,k_0)=P(\sigma|\nu,\lambda)P(\mu|\sigma,\mu_0,k_0)$. \cite{george} specified their priors as
\begin{equation*}
	\sigma^2\sim \frac{\nu\lambda}{\chi_{\nu}^2};\quad\mu|\sigma\sim N(\mu_0,\frac{\sigma^2}{k_0}).
\end{equation*}
$\nu$ is set at 10 to make the spread of error for a single component $k$ tighter. $\lambda$ is chosen using the idea from how $\lambda$ is determined in BART with the quantile set at 0.95 instead of 0.9 (See Appendix \ref{hyperparameters} for how $\lambda$ is determined in BART). For $\mu_0$, because DPMBART subtracts $\bar{{Y}}$ from ${Y}$, $\mu_0=0$. For $k_0$, the residuals of a multiple linear regression fit is used to place $\mu$ into the range of these residuals, ${r}$. The marginal distribution of is $\mu\sim\frac{\sqrt{\lambda}}{\sqrt{k_0}}t_{\nu}$, where $t_{\nu}$ is a t distribution with $\nu$ degrees of freedom. Let $k_s$ be the scaling for $\mu$. Given $k_s=10$, $k_0$ can be chosen by solving 
\[
	\max |r_k|=k_s\frac{\sqrt{\lambda}}{\sqrt{k_0}}.
\]
For $\alpha$, the prior used by DPMBART is the same as in Section 2.5 of \cite{rossi} where the idea is to relate $\alpha$ to the number of unique components in $(a_i,\sigma_i^2)$. 

The posterior draw for DPMBART follows most of the ideas discussed in General BART where first, the idea of Equation (\ref{G_draw}) is used to draw $(T,M)|a_i,\sigma_i^2.$
The slight difference is to view this as a weighted BART draw with $\varepsilon\sim N(0,w_i\sigma^2)$. The second draw,
$(a_i,\sigma^2)|(T,M)$
follows Equation (\ref{H_draw}) which can be solved by using draws (a) and (b) of the algorithm in Section 1.3.3 of Escobar and West in \cite{dey}. The final draw is 
$\alpha|(a_i,\sigma^2)$.
This is obtained by putting $\alpha$ on a grid and the using Bayes' theorem with $P(\alpha|(a_i,\sigma_i^2))=P(\alpha|K)\propto P(K|\alpha)P(\alpha)$ where $K$ is the number of unique $(a_i,\sigma_i^2)$'s.

\section{Discussion}
\label{conclude}
In this tutorial, we walked through the BART model and algorithm in detail, and presented a generalized model based on recent extensions. We believe this is important because of the growing use of BART in research applications as well as being used as a competitor model for new modeling or prediction methods. By clarifying the various components of BART, we hope that researchers will be more comfortable using BART in practice. 

Despite the success of BART, there has been a growing number of papers that point out limitations of BART and propose modifications. One  issue is the inability of BART to do variable selection due to the use of the uniform prior to select the covariate to be split upon in the internal nodes. One simple solution is to allow researchers to place different prior probabilities on each covariate \citep{kapelner_jbart}. Other solutions include  using a Dirichlet Process Prior for selecting covariates \citep{linero} or  using a spike-and-slab prior \citep{liu_abc}. Another commonly addressed issue is the computation speed of BART. Due to the many MH steps that BART require, computation speed of BART can often be slow, especially when the sample size $n$ and/or the number of covriates $p$ is large. One direction is to parallelize the computational steps in BART, which was proposed by \cite{pratola_pbart} and \cite{kapelner_jbart}. The other direction is to improve the efficiency of the MH steps which leads to the reduction in the number of trees needed. Notable examples include \cite{lak}, where particle Gibbs sampling was used to propose the tree structure $T_j$'s; \cite{lisa_bart}, where likelihood inflated sampling was used to calculate the MH steps, and more recently \cite{accel_bart}, where they proposed to use a different tree-growing algorithm which grows the tree from scratch (root node) at each iteration. Other less discussed issues with BART include the problem of under estimation of the uncertainty of BART caused by inefficient mixing when the true variation is small \citep{pratola}, inability of BART to handle smooth functions \citep{yang}, and inclusion of many spurious interactions when the number of covariates is large \citep{du}. Finally, the posterior concentration properties of BART have also been discussed recently by \cite{rockova_bcart}, \cite{rockova_bart}, and \cite{yang}. These works provide theoretical proof of why BART has been successful in many data applications we have seen thus far.

A second component we focused on was how we can extend BART using a very simple idea without having to re-write the whole MCMC algorithm to draw the regression trees.  We term this framework General BART. This framework has already been used by various authors to extend BART to semiparamteric situations where a portion of the model was desired to be linear and more interpretable, correlated outcomes, solve the statistical matching problem in survey, and improve the robustness assumption of the error term in BART. By unifying these methods under a single framework and showing how these methods are related to the General BART model, we hope to provide researchers a guide and inspiration of how to possibly extend BART to their research work where the use of the simple independent continuous or binary BART model is insufficient. For example, researchers working with longitudinal data may want a more flexible modeling portion for the random effects and hence may want to model $H(W,{\Theta})$ as BART. Another possibility is to combine the ideas in Examples \ref{semibart}, \ref{ribart}, and \ref{dpmbart}, i.e. correlated outcomes with an interpretable linear model portion and robust error assumptions. Such are the possibilities for our proposed General BART framework.  

We do note that the critical component of our General BART framework is re-writing the model in such a way that the MCMC draw of the regression trees can be done separately from the rest of the model. In situations where this is not possible, re-writing of the MCMC procedure for the regression trees may be needed. An example of this would occur if, rather than mapping the outcome to a parameter at the terminal node of a regression tree, it is mapped to a regression model. However, we feel that the general BART model is flexible enough to handle many of the extensions that might be of interest to researchers.

\bibliography{references}
\bibliographystyle{biom}

\appendix
\section{Hyperparameters for BART \label{hyperparameters}}
The hyperparameters for continuous outcomes BART that needs to be set are: $\alpha$, $\beta$, $\mu_{\mu}$, $\sigma_{\mu}$, $\nu$, and $\lambda$. These hyperparameters are constructed as a mix of apriori fixed and data-driven. For $\alpha$ and $\beta$, the default values of $\alpha=0.95$ and $\beta=2$ provide a balanced penalizing effect for the probability of a node splitting \citep{chipman_bart}. For $\mu_{\mu}$ and $\sigma_{\mu}$, they are set such that $E[{Y}|X]\sim N(m\mu_{\mu},m\sigma_{\mu}^2)$ assigns high probability to the interval $(\min({Y}),\max({Y}))$. This can be achieved by defining $v$ such that $\min({Y})=m\mu_{\mu}-v\sqrt{m}\sigma_{\mu}$ and $\max({Y})=m\mu_{\mu}+v\sqrt{m}\sigma_{\mu}$. For ease of posterior distribution calculation, ${Y}$ is transformed to become $\tilde{{Y}}=\frac{{Y}-\frac{\min({Y})+\max({Y})}{2}}{\max({Y})-\min({Y})}$. This results in $\tilde{{Y}}\in (-0.5,0.5)$ where $\min({Y})=-0.5$ and $\max({Y})=0.5$. This has the effect of allowing the hyperparamter $\mu_{\mu}$ to be set as 0 and $\sigma_{\mu}$ to be determined as $\sigma_{\mu}=\frac{0.5}{v\sqrt{m}}$ where $v$ is to be chosen. For $v=2$, $N(m\mu_{\mu},m\sigma_{\mu}^2)$ assigns a prior probability of 0.95 to the interval $(\min({Y}),\max({Y}))$ and is the default value. Finally for $\nu$ and $\lambda$, the default value for $\nu$ is 3 and $\lambda$ is the value such that $P(\sigma^2<s^2;\nu,\lambda)=0.9$ where $s^2$ is the estimated variance of the residuals from the multiple linear regression with ${Y}$ as the outcomes and $X$ as the covariates.

For binary outcomes, the $\alpha$ and $\beta$ hyperparameters are the same but the $\mu_{\mu}$ and $\sigma_{\mu}$ hyperparameters are specified differently from continuous outcomes BART. To set the hyperparameters for $\mu_{\mu}$ and $\sigma_{\mu}$, we set $\mu_{\mu}=0$ and $\sigma_{\mu}=\frac{3}{v\sqrt{m}}$ where $v=2$ would result in an approximate 95\% probability that draws of $\sum_{j=1}^m g(X;T_j,{M}_j)$ will be within $(-3,3)$. No transformation of the latent variable ${Z}$ would be needed.

\section{Posterior distributions for $\mu_{kj}$ and $\sigma^2$ in BART}
\subsection{$P(\mu_{kj}|T_j,\sigma,{R}_j)$}

Let ${R}_{kj}=(R_{kj1},\ldots, R_{kjn_k})^T$ be a subset from ${R}_j$ where $n_k$ is the number of $R_{kjh}$s allocated to the terminal node with parameter $\mu_{kj}$ and $h$ indexes the subjects allocated to the terminal node with parameter $\mu_{kj}$. We note that $R_{kjh}|g({X}_{kjh},T_j,{M}_j),\sigma\sim N(\mu_{kj},\sigma^2)$ and $\mu_{kj}|T_j\sim N(\mu_{\mu},\sigma_{\mu}^2)$. Then the posterior distribution of $\mu_{kj}$ is given by
\begin{align*}
	P(\mu_{kj}|T_j,\sigma,{R}_j)&\propto P({R}_{kj}|T_j,\mu_{kj},\sigma)P(\mu_{kj}|T_j)\\
	&\propto \exp[-\frac{\sum_h(R_{kjh}-\mu_{kj})^2}{2\sigma^2}]\exp[-\frac{(\mu_{kj}-\mu_{\mu})^2}{2\sigma_{\mu}^2}]\\
	&\propto \exp[-\frac{(n_k\sigma_{\mu}^2+\sigma^2)\mu_{kj}^2-2(\sigma_{\mu}^2\sum_hR_{kjh}+\sigma^2\mu_{\mu})\mu_{kj}}{2\sigma^2\sigma_{\mu}^2}]\\
	&\propto\exp[-\frac{(\mu_{kj}-\frac{\sigma_{\mu}^2\sum_hR_{kjh}+\sigma^2\mu_{\mu}}{n_k\sigma_{\mu}^2+\sigma^2})^2}{2\frac{\sigma^2\sigma_{\mu}^2}{n_k\sigma_{\mu}^2+\sigma^2}}]
\end{align*}
where $\sum_h(R_{kjh}-\mu_{kj})^2$ is the summation of the squared difference between the parameter $\mu_{kj}$ and the $R_{kjh}$s allocated to the terminal node with parameter $\mu_{kj}$. 

\subsection{$P(\sigma^2|(T_1,{M}_1),\ldots,(T_m,{M}_m),{Y})$}
Let ${Y}=(Y_1,\ldots,Y_n)^T$ and $i$ index the subjects $i=1,\ldots,n$. With $\sigma^2\sim IG(\frac{\nu}{2},\frac{\nu\lambda}{2})$, we obtain the posterior draw of $\sigma$ as follows
\begin{align*}
	P(\sigma^2|(T_1,{M}_1),\ldots,(T_m,{M}_m),{Y})&\propto P({Y}|(T_1,{M}_1),\ldots,(T_m,{M}_m),\sigma)P(\sigma^2)\\
	&= P(Y|\sum_{j=1}^m g(X,T_j,{M}_j),\sigma)P(\sigma^2)\\
	&=\{\prod_{i=1}^n(\sigma^2)^{-\frac{1}{2}}\exp[-\frac{(Y_i-\sum_{j=1}^m g(X_i,T_j,M_j))^2}{2\sigma^2}]\}\\
	&\quad(\sigma^2)^{-(\frac{\nu}{2}+1)}\exp(-\frac{\nu\lambda}{2\sigma^2})\\
	&=(\sigma^2)^{-(\frac{\nu+n}{2}+1)}\\
	&\quad\exp[-\frac{\nu\lambda+\sum_{i=1}^n(Y_i-\sum_{j=1}^m g(X_i,T_j,M_j))^2}{2\sigma^2}]
\end{align*}
where $\sum_j^m g(X_i,T_j,M_j)$ is the predicted value of BART assigned to observed outcome $Y_i$.

\section{Metropolis-Hastings ratio for the grow and prune step}
This section is modified from Appendix A of \cite{kapelner_jbart}. Note that 
\[
	\alpha(T_j,T_j^*)=\min\{1,\frac{q(T_j^*,T_j)}{q(T_j,T_j^*)}\frac{P({R}_j|X,T_j^*,{M}_j)}{P({R}_j|X,T_j,{M}_j)}\frac{P(T_j^*)}{P(T_j)}\}.
\]
where $\frac{q(T_j^*,T_j)}{q(T_j,T_j^*)}$ is the transition ratio, $\frac{P({R}_j|X,T_j^*,{M}_j)}{P({R}_j|X,T_j,{M}_j)}$ is the likelihood ratio, and $\frac{P(T_j^*)}{P(T_j)}$ is the tree structure ratio of Kapelner and Bleich, Appendix A. We now present the explicit formula for each ratio under the grow and prune proposal.

\subsection{Grow proposal}
\subsubsection{Transition ratio}
$q(T_j^*,T_j)$ indicates the probability of moving from $T_j$ to $T_j^*$ i.e. selecting and terminal node and growing two children from $T_j$. Hence,
\begin{align*}
	P(T_j^*|T_j)&=P(grow)P(\text{selecting terminal node to grow from})\times\\
	&\quad P(\text{selecting covariate to split from})\times\\
	&\quad P(\text{selecting value to split on})\\
	&=P(grow)\frac{1}{b_j}\frac{1}{p}\frac{1}{\eta}.
\end{align*}
In the above equation, $P(grow)$ can be decided by the researcher although the default provided is 0.25, $b_j$ is the number of available terminal nodes to split on in $T_j$, $p$ is the number of variables left in the partition of the chosen terminal node, and $\eta$ is the number of unique values left in the chosen variable after adjusting for the parents' splits.

$q(T_j,T_j^*)$ on the other hand indicates a pruning move which involves the probability of selecting the correct internal node to prune on such $T_j^*$ becomes $T_j$. This is given as
\begin{align*}
	P(T_j|T_j^*)&=P(prune)P(\text{selecting the correct internal node to prune})\\
	&=P(prune)\frac{1}{w_2^*}
\end{align*}
where $w_2^*$ denotes the number of internal nodes which have only two children terminal nodes. 

This gives a transition ratio of
\[
	\frac{q(T_j^*,T_j)}{q(T_j,T_j^*)}=\frac{P(T_j^*|T_j)}{P(T_j|T_j^*)}=\frac{P(prune)}{P(grow)}\frac{b_jp\eta}{w_2^*}.
\]
If there are no variables with two or more unique values, this transition ratio will be set to 0.

\subsubsection{Likelihood ratio}
Since the rest of the tree structure will be the same between $T_j^*$ and $T_j$ except for the terminal node where the two children are grown, we need only concentrate on this terminal node. Let $l$ be the selected node and $l_L$ and $l_R$ be the two children of the grow step. Then
\begin{align*}
	\frac{P({R}_j|X,T_j^*,{M}_j)}{P({R}_j|X,T_j,{M}_j)}&=\frac{P({R}_{l_{(L,1)},j},\ldots,{R}_{l_{(L,n_L)},j}|\sigma^2)P({R}_{l_{(R,1)},j},\ldots,{R}_{l_{(R,n_R)},j}|\sigma^2)}{P({R}_{1,j},\ldots,{R}_{n_l,j}|\sigma^2)}\\
	&=\sqrt{\frac{\sigma^2(\sigma^2+n_l\sigma_{\mu}^2)}{(\sigma^2+n_L\sigma_{\mu}^2)(\sigma^2+n_R\sigma_{\mu}^2)}}\exp[\frac{\sigma_{\mu}^2}{2\sigma^2}(\frac{(\sum_{k=1}^{n_L}{R}_{l_{(L,k)},j})^2}{\sigma^2+n_L\sigma_{\mu}^2}\\
	&\quad+\frac{(\sum_{k=1}^{n_R}{R}_{l_{(R,k)},j})^2}{\sigma^2+n_R\sigma_{\mu}^2}-\frac{(\sum_{k=1}^{n_l}{R}_{l_{(l,k)},j})^2}{\sigma^2+n_l\sigma_{\mu}^2})].
\end{align*}

\subsubsection{Tree structure ratio}
Because the $T_j$ can be specified using three aspects, we let $P_{SPLIT}(\theta)$ denote the probability that a selected node $\theta$ will split and $P_{RULE}(\theta)$ denote the probability that which variable and value is selected. Then based on $P_{SPLIT}(\theta)\propto \frac{\alpha}{(1+d_{\theta})^{\beta}}$ and because $T_j$ and $T_j^*$ only differs at the children nodes, we have
\begin{align*}
	\frac{P(T_j^*)}{P(T_j)}&=\frac{\prod_{\theta\in H_{terminals}^*}(1-P_{SPLIT}(\theta))\prod_{\theta\in H_{internals}^*}P_{SPLIT}(\theta)\prod_{\theta\in H_{internals}^*}P_{RULE}(\theta)}{\prod_{\theta\in H_{terminals}}(1-P_{SPLIT}(\theta))\prod_{\theta\in H_{internals}}P_{SPLIT}(\theta)\prod_{\theta\in H_{internals}}P_{RULE}(\theta)}\\
	&=\frac{[1-P_{SPLIT}(\theta_L)][1-P_{SPLIT}(\theta_R)]P_{SPLIT}(\theta)P_{RULE}(\theta)}{1-P_{SPLIT}(\theta)}\\
	&=\frac{(1-\frac{\alpha}{(1+d_{\theta_L})^{\beta}})(1-\frac{\alpha}{(1+d_{\theta_R})^{\beta}})\frac{\alpha}{(1+d_{\theta})^{\beta}}\frac{1}{p}\frac{1}{\eta}}{\frac{\alpha}{(1+d_{\theta})^{\beta}}}\\
	&=\alpha\frac{(1-\frac{\alpha}{(2+d_{\theta})^{\beta}})^2}{[(1+d_{\theta})^{\beta}-\alpha]p\eta}
\end{align*}
because $d_{\theta_L}=d_{\theta_R}=d_{\theta}+1$.

\subsection{Prune proposal}
Since prune is the direct opposite of the grow proposal, the explicit formula of $\alpha(T_j,T_j^*)$ will just be the inverse of the grow proposal.

\end{document}